\documentclass[10pt,letterpaper]{article}
\usepackage{opex3}

\usepackage{graphicx}
\usepackage{epstopdf}

\usepackage{cite}

\usepackage{amsmath}

\usepackage{amsfonts}

\usepackage[colorlinks,urlcolor=blue,linkcolor=blue,citecolor=blue,anchorcolor=blue]{hyperref}

\begin{document}

\title{Azimuthons in weakly nonlinear waveguides of different symmetries}

\author{Yiqi Zhang$^{1,2}$, Stefan Skupin$^{1,3}$, Fabian Maucher$^1$, Arpa Galestian Pour$^3$, Keqing Lu$^{2}$, Wieslaw Kr\'olikowski$^4$}

\address{$^1$Max Planck Institute for the Physics of Complex Systems, N\"{o}thnitzer Stra{\ss}e 38, 01187 Dresden, Germany\\
$^2$State Key Laboratory of Transient Optics and Photonics, Xi'an Institute of Optics and Precision Mechanics of Chinese Academy of Sciences, 710119 Xi'an, China\\
$^3$Institute of Condensed Matter Theory and Solid State Optics, Friedrich Schiller University, Max-Wien-Platz 1, 07743 Jena, Germany\\
$^4$Laser Physics Center, Research School of Physics and Engineering, Australian National University, Canberra, ACT 0200, Australia}




\begin{abstract}
We show that weakly guiding nonlinear waveguides support stable
propagation of rotating spatial solitons (azimuthons). We investigate the role of waveguide symmetry on
the soliton rotation.
We find that azimuthons in circular waveguides  always rotate rigidly during propagation
and the analytically predicted rotation frequency is in excellent agreement with numerical simulations.
On the other hand, azimuthons in square waveguides may experience spatial  deformation during propagation.
Moreover, we show that there is  a critical value for the modulation depth of azimuthons
 above which solitons just wobble back and forth, and below which they rotate continuously.
 We explain these dynamics using the concept of energy difference between different orientations of the azimuthon.
\end{abstract}



\section{Introduction}

Spatial solitons are nonlinear localized states
that keep their form during propagation due to the balance between diffraction and self-induced nonlinear potential \cite{Stegeman_science_1999}.
Recently, there has been a lot of interest in a generalized type of spatial solitons, the so-called azimuthons.
These are azimuthally modulated  beams, that exhibit
steady angular rotation upon propagation~\cite{desyatnikov_prl_2005}.
They can be considered as azimuthally perturbed optical vortices,
i.e. beams with singular phase structure~\cite{coullet_oc_1989, desyatnikov_prl_2005, lashkin_pra_2008}.
Theoretical studies demonstrated both, stable and unstable propagation of azimuthons \cite{buccoliero_ol_2008, lopez-aguayo_oe_2006, buccoliero_prl_2007},
and the first experimental observation of optical azimuthons  was recently achieved in rubidium vapors~\cite{minovich_oe_2009}.

It appears that higher order solitonic structures and optical vortices are generally
unstable in typical nonlinear media with local (Kerr-like) response~\cite{kruglov_jmo_1992, skryabin_pre_1998}.
On the other hand, it has been shown that a spatially nonlocal nonlinear response provides stabilization of various complex solitonic  structures including vortices~\cite{suter_pra_1993, bang_pre_2002, krolikowski_job_2004, briedis_oe_2005}. Consequently,
azimuthons and their dynamics have been studied almost exclusively in the context of
spatially nonlocal nonlinear media~\cite{lopez-aguayo_ol_2006, buccoliero_prl_2007, stefan_oe_2008, Fabian:oqe:09}.
In spite of the fact that there are various physical settings exhibiting nonlocality such as nematic liquid crystals \cite{peccianti_ol_2002, conti_prl_2003, conti_prl_2004},
Bose-Einstein condensates \cite{lashkin_pra_2008, nath_pra_2007, lashkin_pra_2009}, plasmas \cite{litvak_plasmas_1975}, thermo-optical materials \cite{rotschild_prl_2005} etc.,
experimental realization of such systems is always quite involved. Moreover, from the theoretical point of view, nonlocal media are quite challenging for numerical modeling and analytical treatment.

In this paper we propose a much simpler and experimentally accessible
optical system to study the propagation of azimuthons: a weakly nonlinear optical multi-mode waveguide.
Here, weakly nonlinear means that the nonlinear induced index change, which is proportional to the intensity of the optical beam, is small
compared to the index profile (or trapping  potential) of the waveguide.
Following \cite{stefan_pre_2004}, we can expect that weakly nonlinear azimuthons are stable in multi-mode waveguides.

The paper is organized as follows: In Sec.~\ref{modelling},
we briefly introduce the general model equation for beam propagation in weakly nonlinear waveguides,
and then we discuss in detail the properties of dipole azimuthons in circular and square waveguides
in Sec.~\ref{dipoles}, respectively.
In Sec.~\ref{higherorder}, higher order azimuthons are investigated, and in Sec.~\ref{conclusion} we conclude.

\section{Mathematical modeling}
\label{modelling}
We consider the evolution of a continuous wave (CW) optical beam with amplitude $\mathcal{E} (\xi, \eta, \zeta)$, where
$(\xi,\eta)$ and $\zeta$ denote the transverse and longitudinal coordinates, respectively.
Then the propagation of this beam 
 in a weakly-guiding waveguide with Kerr nonlinearity in the scalar,
slowly varying envelope approximation is described by the following equation \cite{boyd_book_2008}:
\begin{equation}
 i \frac{\partial}{\partial \zeta} \mathcal{E} +
\frac{1}{2k_0} \left( \frac{\partial ^2}{\partial \xi^2} + \frac{\partial ^2}{\partial \eta^2} \right) \mathcal{E} +
k_0 \frac{n_2}{n_b} |\mathcal{E}|^2 \mathcal{E} + k_0 \frac{n(\xi,\eta)-n_b}{n_b} \mathcal{E}=0,
\label{physicaleq}
\end{equation}
where $k_0=2 \pi n_b / \lambda_0$ refers to the carrier central wave number in the medium; $n_2$ is the Kerr nonlinear coefficient,
$n(\xi,\eta)$ the linear refractive index distribution,
$n_b$ the background index, and $\lim_{\xi,\eta \rightarrow \infty}n(\xi,\eta)=n_b$. We consider a weakly-guiding
waveguide, so both the linear and nonlinear induced index change $|n-n_b|$ and $n_2|\mathcal{E}|^2$ are small compared to
the mean index $n_b$, and at the same time $n_2|\mathcal{E}|^2 \ll |n-n_b|$, to guarantee a weak nonlinearity.

From the mathematical point of view, Eq.~(\ref{physicaleq}) is a (2+1)-dimensional nonlinear Schr{\"o}dinger (NLS) equation
with linear  potential representing the waveguide profile.
For technical convenience, we rescale Eq.~(\ref{physicaleq}) to dimensionless quantities with $x=\xi/r_0$, $y=\eta/r_0$,
$z=\zeta/(2 k_0 r_0^2)$, and $\sigma=\mathrm{sgn}(n_2)$, where $r_0$ represents the transverse spatial extent of the waveguide.
Then, the two-dimensional (2D) NLS equation for the scaled wave function $\psi = k_0 r_0 \sqrt{2|n_2|/n_b} \mathcal{E}$ reads
\begin{equation}
i\frac{\partial}{\partial z} \psi + \left( \frac{\partial ^2}{\partial x^2}
 + \frac{\partial ^2}{\partial y^2} \right) \psi +
\sigma |\psi|^2 \psi + V \psi=0,
\label{partial}
\end{equation}
with an ``attractive'' bounded potential $V=2k_0^2 r_0^2 (n-n_b)/n_b$, given by the spatial profile of the refractive  index of the waveguide. Here  we will consider propagation in step index waveguides with parameter $V(x,y)$  given as follows
\begin{equation*}
V(x,y)=
\begin{cases}
V_0 & \mathrm{~where~}
\begin{cases}
\sqrt{x^2+y^2} \leq 1 & \mathrm{~for~circular~waveguide}
\\
|x| \leq 1 ~\&~ |y| \leq 1 & \mathrm{~for~square~waveguide}
\end{cases}
\\
0 & \mathrm{elsewhere}
\end{cases}.
\end{equation*}
$V_0$ is the height of the potential and determines the number
of linear waveguide modes.

Considering the specific case of silica-made waveguides, typical values for the parameters are $n_b=1.4$, $|n-n_b| \leq 9 \times 10^{-3}$,
$n_2 = 3\times 10^{-16} ~\mathrm{cm}^2/\mathrm{W}$ at a vacuum wavelength of $\lambda_0=790$~nm, and those values do not vary much when we 
choose any $\lambda_0$ in the range from visible to near-infrared. The laser-induced-damage-threshold
(LIDT) for synthetic fused silica is about $10 ~\mathrm{J}/\mathrm{cm}^2$ for nanosecond pulses (according to Eq.~(1) and Table 2
in Ref.~\cite{kuzuu_ao_1999}) corresponding to $\sim 10^{10} ~\mathrm{W}/\mathrm{cm}^2$, up to which the model should be valid.
For shorter, femtosecond pulses this damage threshold is $1000$ times higher,
but since we consider ``CW'' beams the lower value for nanosecond pulses is relevant.
It is justified to neglect temporal effects on beam propagation,
because the dispersion length is already of the order of kilometers for pulse durations of a few tens of picoseconds,
much longer than propagation distances considered in this work.
Throughout this paper, we choose $V_0=1000$,
and find  that $r_0 \approx 25.0 ~\mu \mathrm{ m}$ through $r_0=\sqrt{\left|n_bV/\left[2k_0^2(n-n_b)\right]\right|}$,
which nicely corresponds to  the radius of a standard circular multi-mode fiber~\cite{agrawal_book_2009}.
In addition, in order to guarantee intensities smaller than the LIDT the condition
$|\psi|^2<1/3$ should be fulfilled.

Note that in Eq.~(\ref{partial}) $\sigma$ is the sign of nonlinearity:
$\sigma =1$ ($\sigma=-1$) represents focusing (defocusing)  nonlinearity.
The main difference between weakly nonlinear waveguides with
focusing and defocusing nonlinearity is that defocusing nonlinearity supports higher amplitudes of the wave function,
and in general leads to more stable and robust configurations. In this paper, however, we consider the experimentally relevant case of a focusing nonlinearity.

\section{Rotating localized dipoles}
\label{dipoles}
\subsection{Dipole azimuthons in circular waveguides}
Azimuthons are a straightforward generalization of the usual ansatz for stationary solutions \cite{desyatnikov_prl_2005}.
They represent spatially rotating structures and hence involve an additional parameter,
the rotation frequency $\omega$ (see also \cite{skryabin_pre_2002}),
so we seek approximate solutions of the form
\begin{equation}
\psi (r,\phi,z) = U (r, \phi - \omega z) \exp (i \kappa z),
\label{ansatz}
\end{equation}
where $r=\sqrt{x^2+y^2}$ and $\phi$ the azimuthal angle in the transverse plane $(x,y)$, $U$ is the stationary profile,
$\omega$ the rotation frequency, and $\kappa$ the propagation constant. For $\omega=0$, azimuthons become ordinary (non-rotating) solitons.
The simplest family of azimuthons is the one connecting the dipole soliton with the single charged vortex soliton \cite{lopez-aguayo_ol_2006}.
A single charged vortex consists of two equal-amplitude dipole-shaped structures representing real and imaginary part of $U$.
If these two components differ in amplitude, the resulting structure forms a ``rotating dipole'' azimuthon.
If one of the components is zero we deal with the dipole soliton, which consists of two out-of-phase humps and does not rotate for symmetry reasons.
In a first attempt,
let us assume we know the radial shape of the linear vortex mode $F(r)$ which is normalized according to $\pi \int r|F(r)|^2 \mathrm{d}r =1$.
Then, using separation of variables, we consider the simplest so-called ``rotating dipole'' azimuthon with ansatz~\cite{stefan_oe_2008}
\begin{equation}
U(r,\phi-\omega z)=AF(r)\left[\cos(\phi-\omega z)+i B \sin(\phi-\omega z) \right],
\label{complexamplitude}
\end{equation}
where $A$ is an amplitude factor, and $1-B$ the azimuthal modulation depth of the resulting ring-like structure.
Because we are operating in the weakly nonlinear regime, using linear waveguide modes as initial conditions for nonlinear (soliton) solutions is a quite good approximation.

After plugging Eq.~(\ref{ansatz}) into Eq.~(\ref{partial}), multiplying by $U^*$ and $\partial U^* / \partial \phi$ respectively,
and integrating over the transverse coordinates we end up with~\cite{stefan_oe_2008}
\begin{subequations}
\begin{equation}
-\kappa P + \omega L_z + I + N = 0,
\end{equation}
\begin{equation}
-\kappa L_z + \omega P' + I' + N' = 0.
\end{equation}
\end{subequations}
This system relates the propagation constant $\kappa$ and the rotation frequency $\omega$ of the azimuthons to integrals over their stationary amplitude profiles:
\begin{subequations}
\begin{equation}
P = \iint r|U(r)|^2 ~\mathrm{d}r~\mathrm{d}\phi,
\end{equation}
\begin{equation}
L_z = -i\iint r\frac{\partial U(r)}{\partial \phi} U^*(r) ~\mathrm{d}r~\mathrm{d}\phi,
\end{equation}
\begin{equation}
I = \iint rU^*(r) \Delta_{\perp} U(r) ~\mathrm{d}r~\mathrm{d}\phi,
\end{equation}
\begin{equation}
N = \iint r\left[\sigma|U(r)|^2+V \right] |U(r)|^2 ~\mathrm{d}r~\mathrm{d}\phi,
\end{equation}
\begin{equation}
P' = \iint r\left| \frac{\partial U(r)}{\partial \phi} \right|^2 ~\mathrm{d}r~\mathrm{d}\phi,
\end{equation}
\begin{equation}
I' = i\iint r\frac{\partial U^*(r)}{\partial \phi} \Delta_{\perp} U(r) ~\mathrm{d}r~\mathrm{d}\phi,
\end{equation}
\begin{equation}
N' = i\iint r\left[\sigma|U(r)|^2+V \right] \frac{\partial U^*(r)}{\partial \phi} U(r) ~\mathrm{d}r~\mathrm{d}\phi.
\end{equation}
\end{subequations}
The first two quantities $(P$ and $L_z)$ have straightforward physical meanings, namely power and angular momentum.
The integrals $I$ and $I'$ are related to the diffraction mechanism of the system, 
whereas $N$ and $N'$ account for waveguide and nonlinearity.
Thus we can formally solve for the rotation frequency with these quantities and obtain
\begin{equation}
\omega=\frac{P(I'+N')-L_z(I+N)}{L_z^2-PP'}.
\label{angularfrequency}
\end{equation}

After inserting Eq.~(\ref{complexamplitude}) into Eq.~(\ref{angularfrequency}),
it turns out that only the nonlinear term contributes to the rotation frequency $\omega$ (see also Eq.~(10) in Ref.~\cite{fabian_pra_2010}).
In order to give an estimate for the rotation frequency, 
we use  the linear stationary modal profile $F(r)$ of a circular waveguide expressed in terms of 
Bessel functions of first kind $J_1$ and modified Bessel function of second kind $K_1$:
\begin{equation}
F(r) = C \times
\begin{cases}
J_1 ( \sqrt{V_0-\kappa} r) & \mathrm{for}~ 0\leq r\leq1
 \\
\dfrac{J_1(\sqrt{V_0-\kappa})}{K_1(\sqrt{\kappa})} \cdot K_1 (\sqrt{\kappa}r) & \mathrm{for}~ r>1
\end{cases},
 \end{equation}
where $C$ is a normalization factor such that $\pi\int|F(r)|^2rdr=1$.
Thus, the analytically predicted rotation frequency is
\begin{equation}
\omega=\frac{\pi \sigma}{2} \int r |F(r)|^4 \mathrm{d}r \cdot A^2 B.
\label{analyticalangular}
\end{equation}
From the above equation,  one can see that sense of rotation is opposite for focusing and defocusing nonlinearities.
Moreover, as $\omega = 2 k_0 r_0^2 \cdot 2 \pi/ \ell$, one can find from Eq.~(\ref{analyticalangular}) an expression for the physical distance $\ell$ over which the azimuthon in a circular waveguide performs one full rotation:
\begin{equation}
\ell= \frac{\displaystyle 8 k_0 r_0^2}{\displaystyle A^2 B \int r |F(r)|^4 \mathrm{d}r}.
\label{physicaldistance}
\end{equation}
For example,
assuming propagation in a silica fiber with the parameters discussed earlier (see Sec.~\ref{modelling}) and taking
$A=0.4$, $B=0.5$,
we obtain $\ell \approx 2.4 ~\mathrm{m}$.

\begin{figure}[htbp]
\centering
\includegraphics[width=0.4\textwidth]{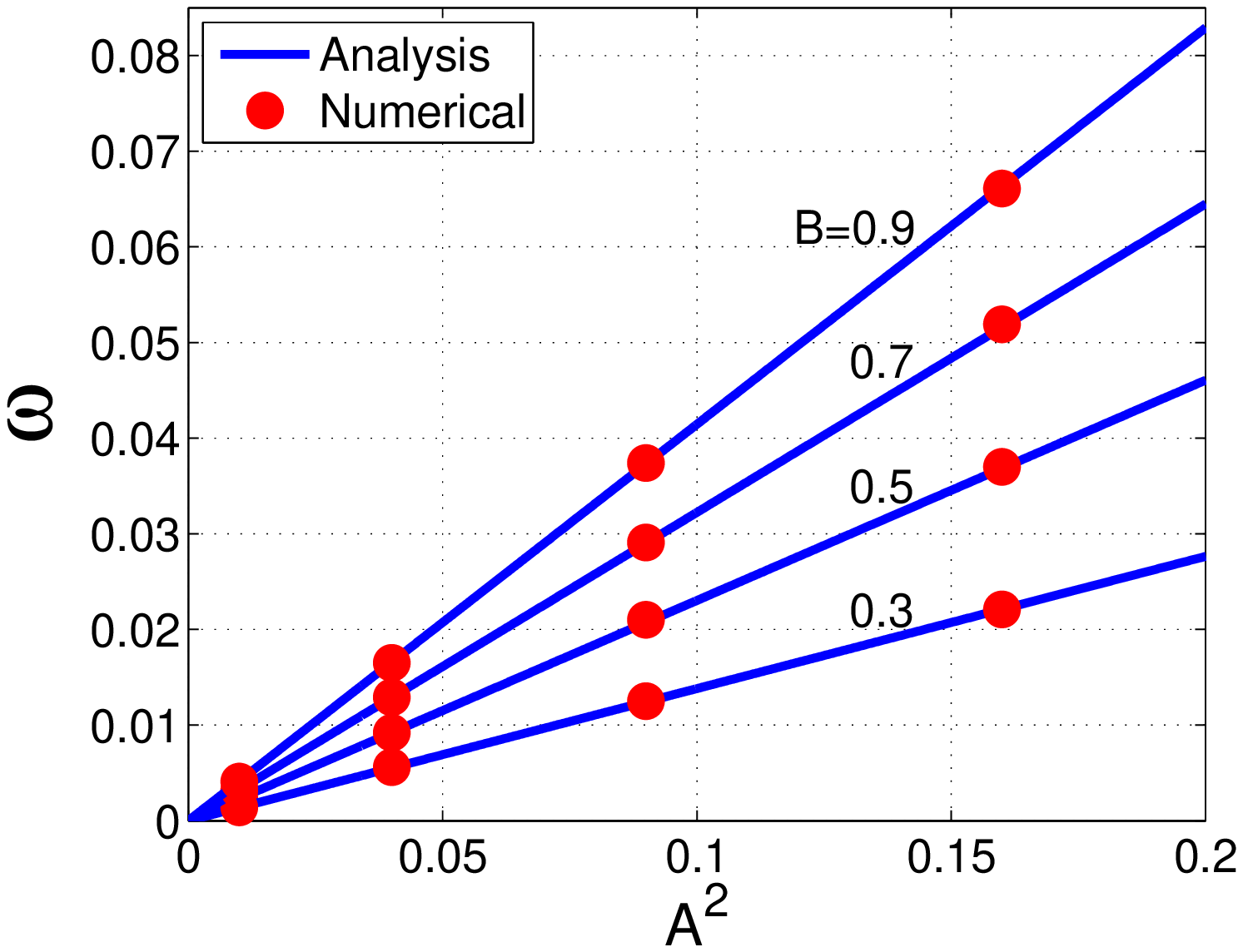} \quad
\includegraphics[width=0.4\textwidth]{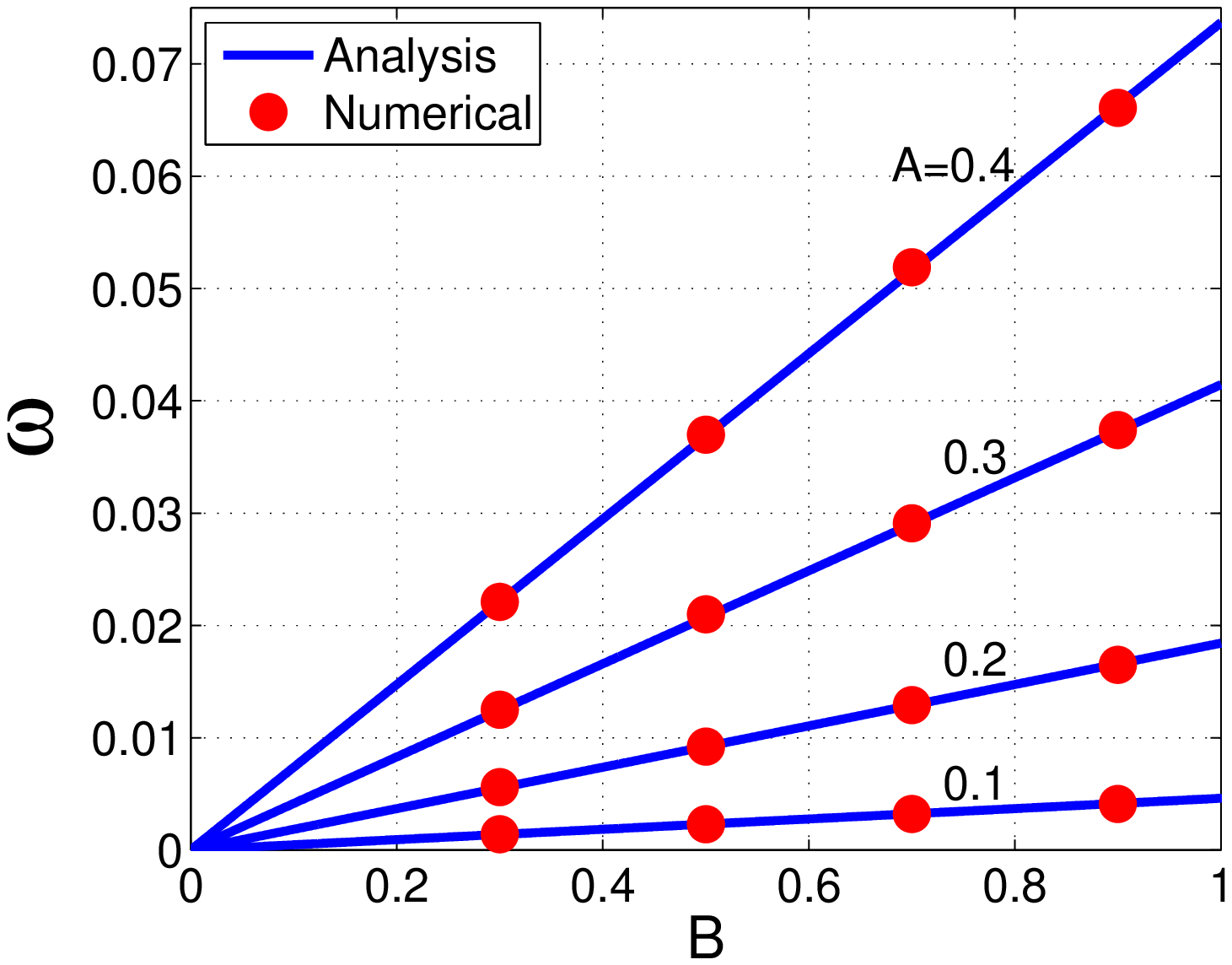}
\caption{Azimuthon rotation frequency $\omega$ versus amplitude factor $A$ (left panel) and amplitude ratio $B$ (right panel).
Blue solid lines show analytical predictions from Eq.~(\ref{analyticalangular}).
Red dots denote results obtained from numerical simulations of Eq.~(\ref{partial}).
The values of $A$ and $B$ are shown next to the lines.}
\label{circularrelations}
\end{figure}

\begin{figure}[htbp]
\centering
\includegraphics[width=0.55\textwidth]{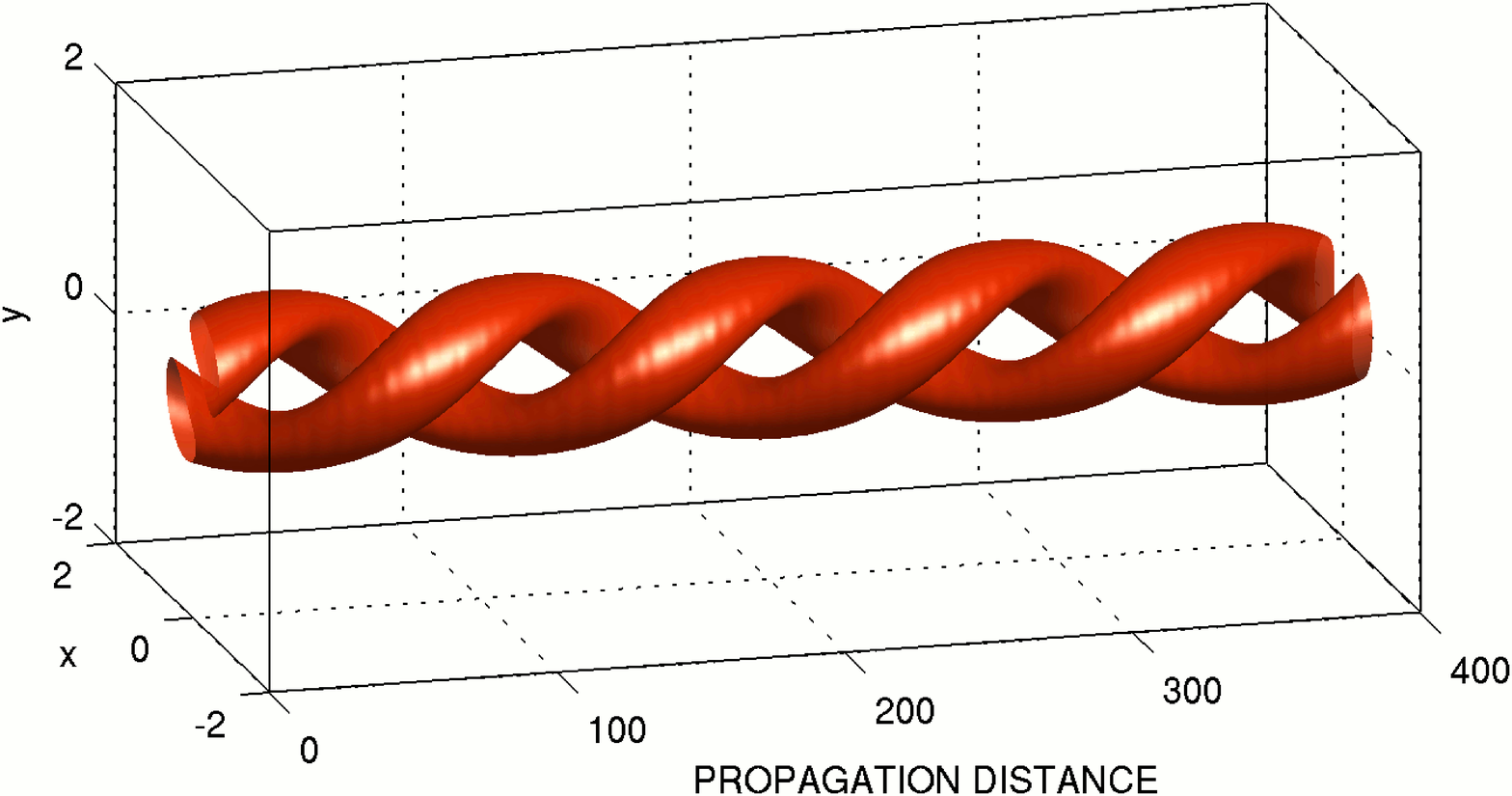} \\
\vspace{0.4cm}
\includegraphics[width=0.35\textwidth]{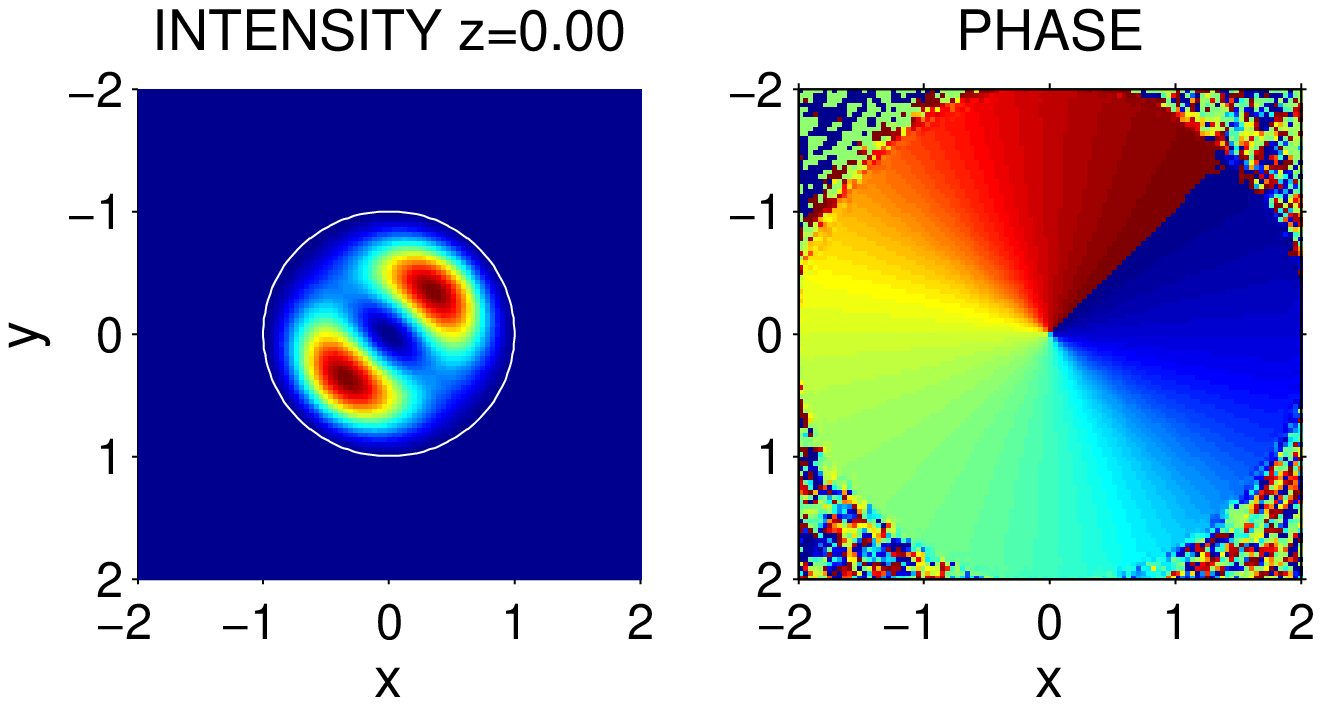} \quad
\includegraphics[width=0.35\textwidth]{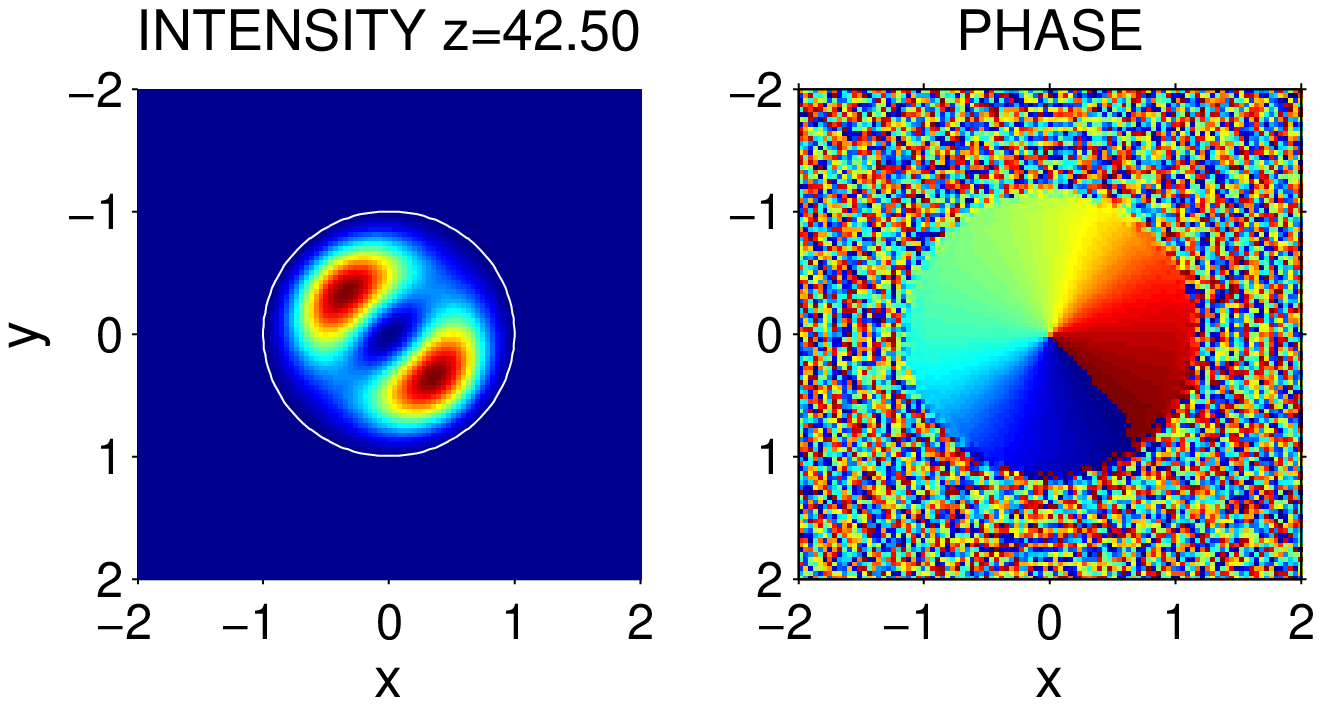}
\caption{The propagation of a dipole azimuthon with $A=0.4,~B=0.5$ in the circular waveguide.
The iso-surface plot on the top clearly displays the spiraling  of the azimuthon during propagation.
Figures in the row below depict the azimuthon's transverse intensity distribution at input, and after rotating $\pi/2$, respectively.
The white line indicates the waveguide boundaries.}
\label{circularevolution}
\end{figure}

Figure~\ref{circularrelations} shows the dependence of the azimuthon
rotation frequency as a function of its  amplitude  $A$ (left panel)
and the modulation parameter $B$ (right panel).
Blue solid lines represent the analytical predictions and, in excellent agreement, red dots are obtained from numerical simulations.
An exemplary propagation of the dipole azimuthon in circular waveguide rotating with  $\omega \approx 0.037$
is illustrated in Fig.~\ref{circularevolution}.
The top iso-surface plot displays 3D rigid and continuous rotation of the azimuthon during propagation.
The row below depicts
the transverse intensity (left)  and  phase (right) distribution of the dipole azimuthon
during propagation after rotating by $\pi/2$.

\subsection{Azimuthon-like dipoles in square waveguides}
\begin{figure}[htbp]
\centering
\includegraphics[width=0.35\textwidth]{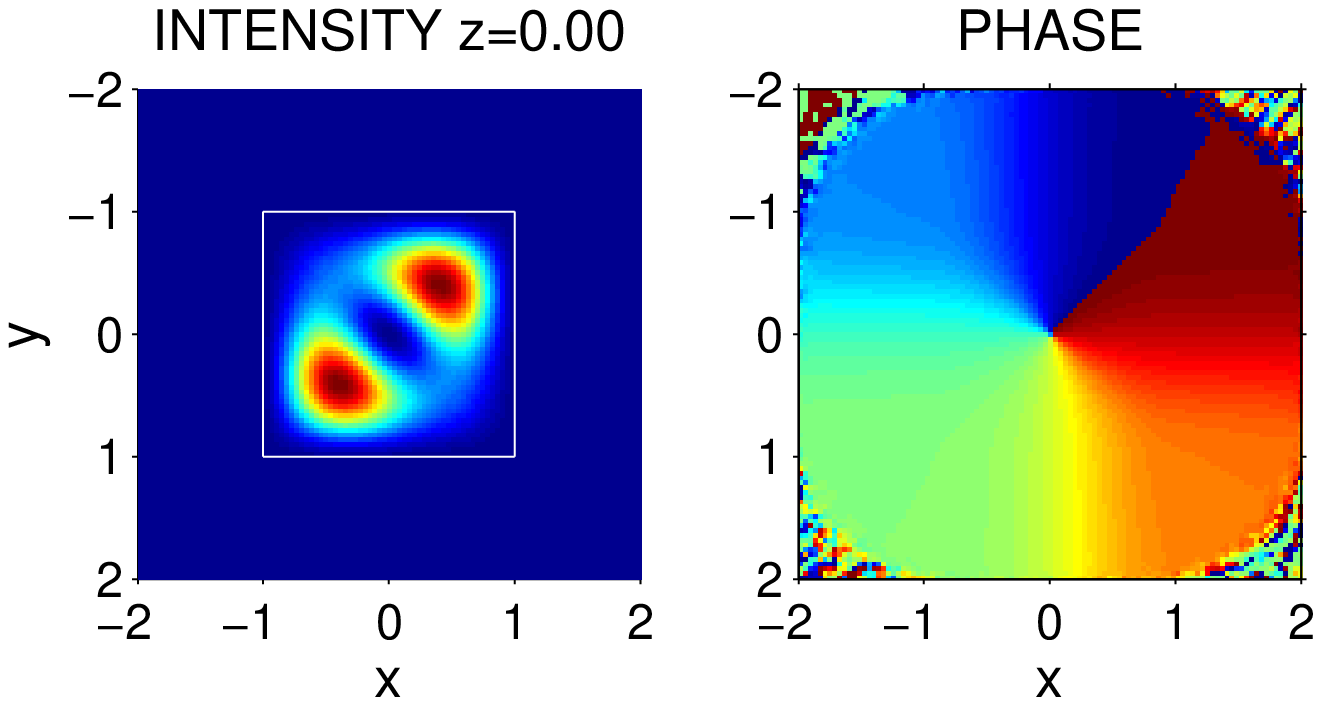} \quad
\includegraphics[width=0.35\textwidth]{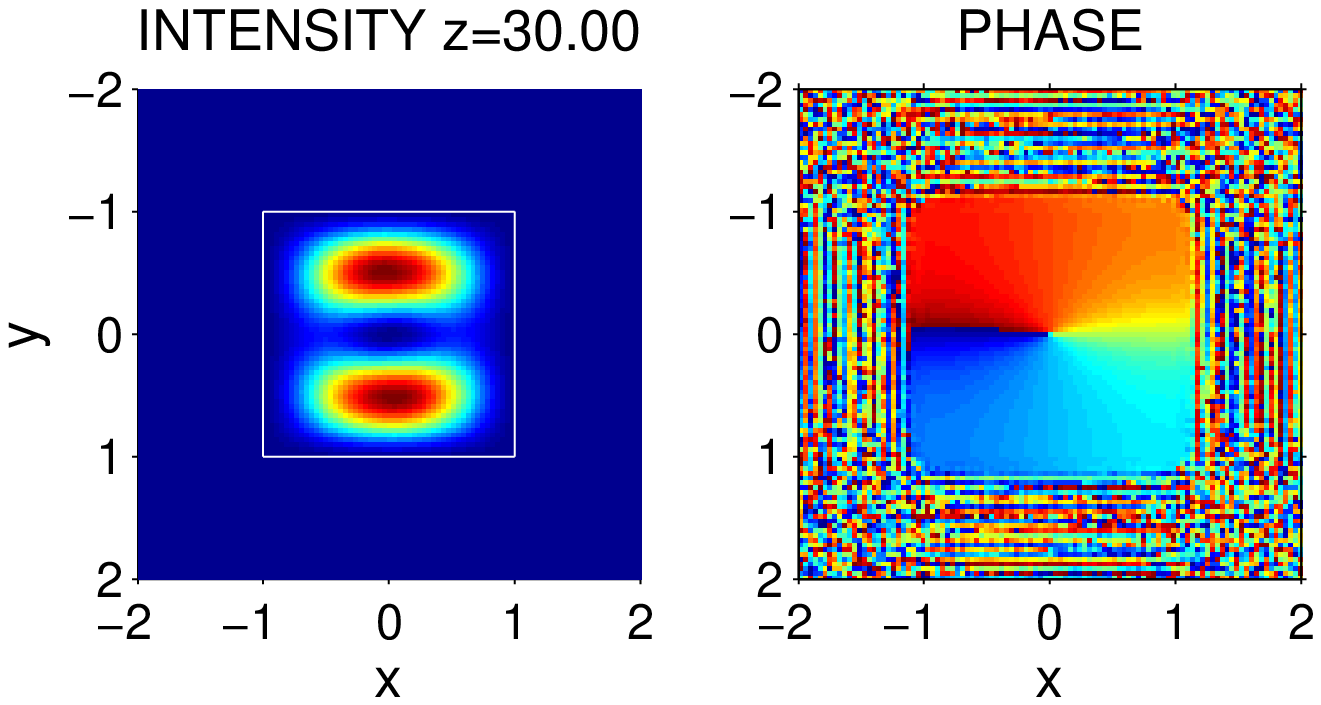} \\
\includegraphics[width=0.55\textwidth]{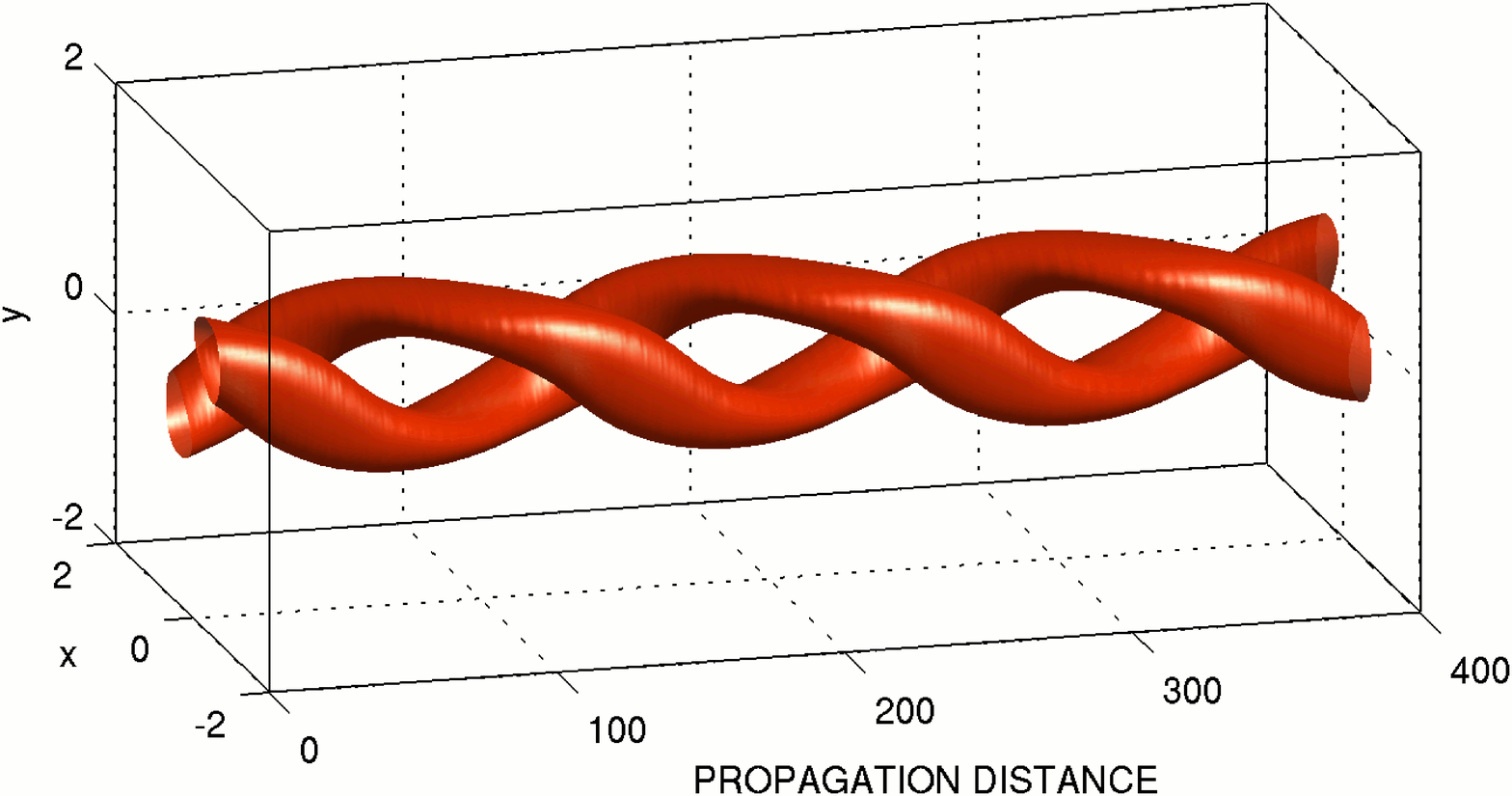}
\caption{The propagation of the azimuthon-like dipole, that rotates in the square waveguide.
Top row: intensity (left) and phase (right) distribution corresponding to the input and soliton rotation by 
 $\pi/4$, respectively. The white line indicates the waveguide boundaries.
The iso-surface plot at the bottom displays the rotation and deformation during propagation.
The initial amplitude  and modulation parameters  are  $A=0.4,~B=0.5$.}
\label{square_r_evolution}
\end{figure}

Because of the lack of circular symmetry,  azimuthons in the strict sense of Eq.~(\ref{complexamplitude}) (preservation of shape during rotation with constant frequency) cannot exist in a square waveguide.
However, as we will show below,  the azimuthon-like behavior is still possible even though the beam propagation is accompanied by variation of the beam transverse intensity distribution.
To set the initial condition, we use two linear degenerated orthogonal dipole modes $D_1$, $D_2$  (as in the case of the circular waveguide),
which are normalized according to $\iint |D_{1,2}|^2~\mathrm{d}x~\mathrm{d}y=1$, and
superpose them as before to form the azimuthon-like object
\begin{equation}
U(x,y,z=0)=A\left(D_1+iBD_2\right).
\label{squareansatz}
\end{equation}
The field Eq.~(\ref{squareansatz}) is then used as an initial condition to the nonlinear Schr{\"o}dinger  equation
Eq.~(\ref{partial}). However, in contrast to the circular waveguide, the orientation of the dipoles $D_1$ and $D_2$ is important
in nonlinear regime. 
If the two orthogonal dipoles $D_1$ and $D_2$ are oriented along the diagonals of the waveguide cross-section (see first subplot of Fig.~\ref{square_r_evolution}),
for a given amplitude $A$ rotation occurs only if the modulation parameter $B$ exceeds  a critical value  $B=B_{\rm cr}$.
Moreover, the rotation is no longer constant  as in the case of cylindrical waveguide but fluctuates and hence in what follows  we use its average value (termed ``average frequency'')  $\bar \omega=1/L\int_0^L\omega(z)~\mathrm{d}z$ with $L$ being the propagation distance corresponding to one full $2\pi$ rotation.
This threshold value $B_{\rm cr}$ decreases if we choose different initial dipole orientations \footnote{In the linear system, any dipole orientation is possible, 
and can be constructed from orthogonal basis ($D_1,D_2$) by superposition.}, and appears to be zero (within our numerical accuracy) 
for parallel orientation with respect to the waveguide boundaries.

\begin{figure}[htbp]
\centering
\includegraphics[width=0.35\textwidth]{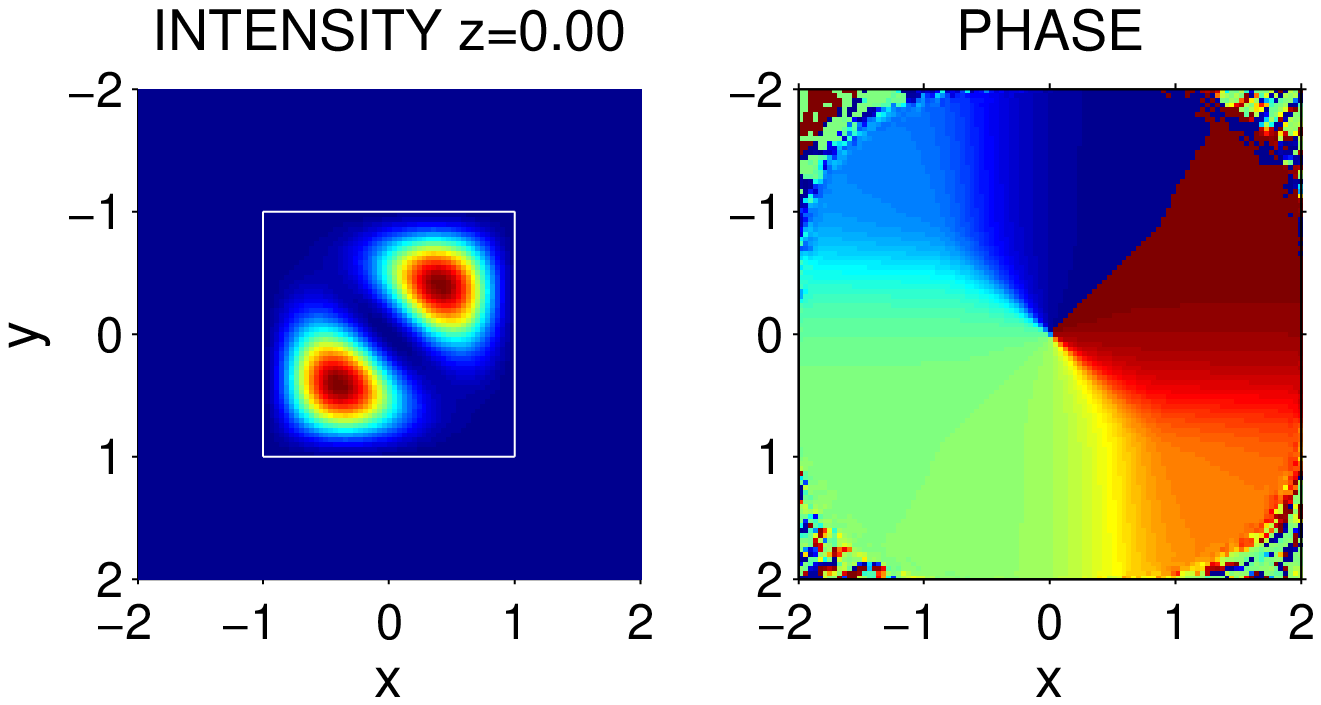} \quad
\includegraphics[width=0.35\textwidth]{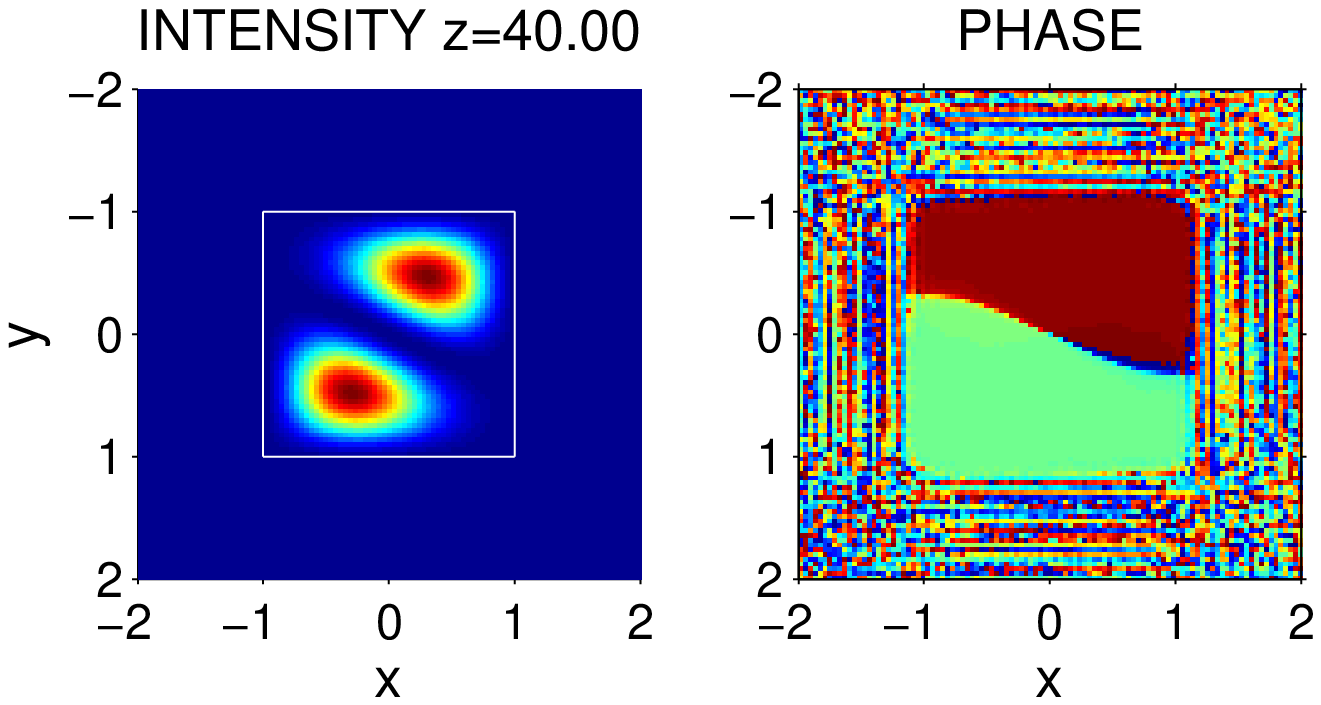} \\
\includegraphics[width=0.55\textwidth]{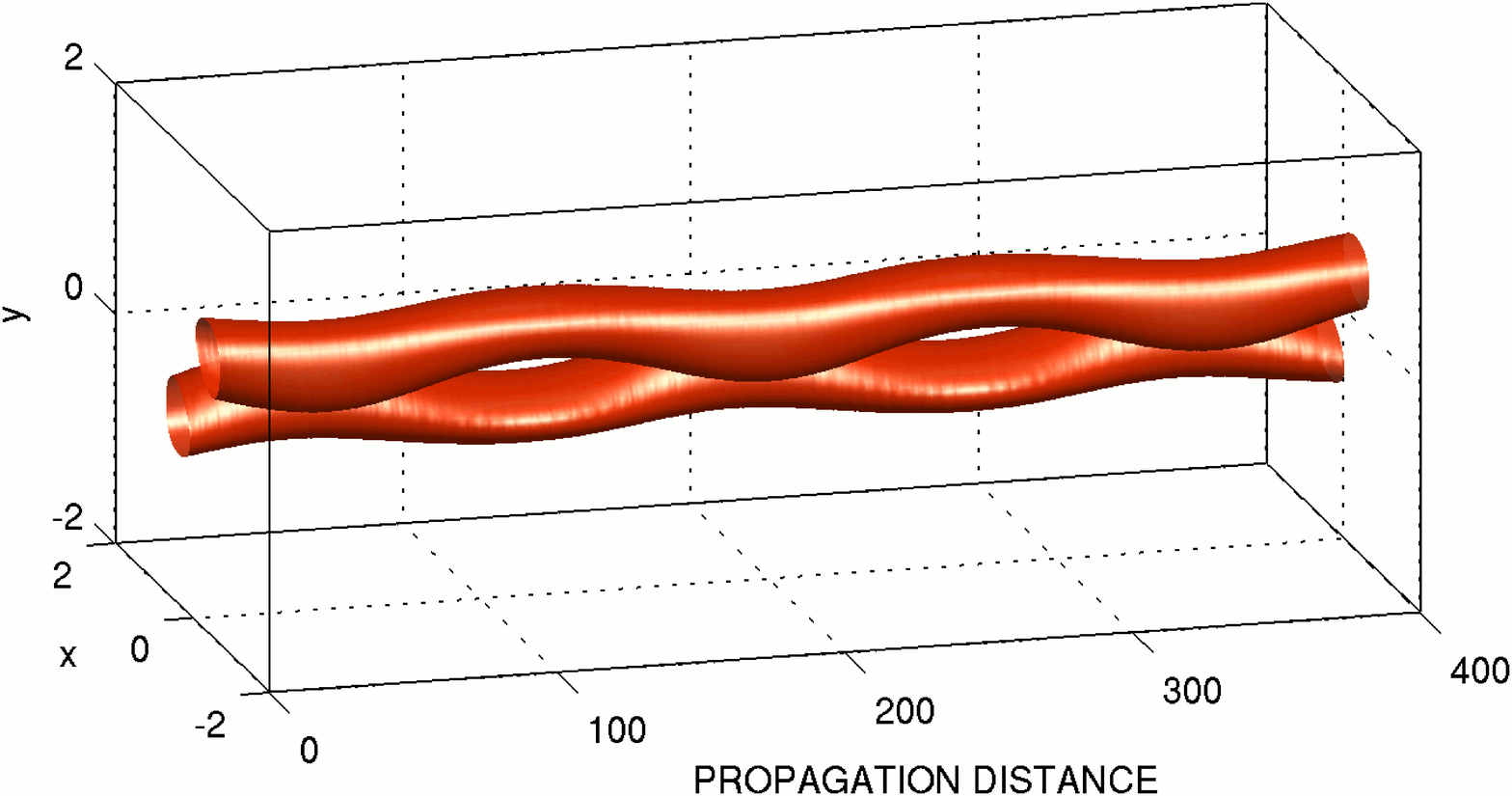}
\caption{The propagation of a dipole azimuthon that wobbles in the square waveguide.
The first row depicts the dipole at input and maximum displacement, respectively. 
The white line indicates the waveguide boundaries. 
The iso-surface plot below displays the twist and deformation during propagation.
The chosen amplitude factor and ratio are $A=0.4,~B=0.2$.}
\label{square_t_evolution}
\end{figure}

Let us focus on the case where the initial dipole-like field structure is 
oriented along the diagonal of the waveguide cross-section, and from now on the notation $D_1$ and $D_2$ stands for this orientation.
Figure~\ref{square_r_evolution} shows an exemplary propagation of the resulting azimuthon-like solution with $B>B_{\rm cr}$,
and an average rotation frequency of  $\bar \omega \approx 0.0262$. The top row depicts intensity and phase distribution at different propagation distance while the bottom plots illustrate the full 3D evolution of the ``soliton''. The rotation (counter-rotating w.r.t.\ phase) accompanied by beam deformation is clearly visible. 
For an amplitude ratio $B$ less than the critical value ($B<B_{\rm cr}$) the azimuthon no longer rotates but  swings back and forth  and hence its average rotation frequency is zero.
The propagation of this wobbling dipole with an amplitude ratio $B$ smaller than the critical value is displayed in Fig.~\ref{square_t_evolution}. 
The right panels show the maximum angle  which the dipole attains  during propagation.
The 3D surface plot in the bottom row  clearly illustrates the swinging movement of the ``soliton'' upon propagation. 

\begin{figure}[htbp]
\centering
\includegraphics[width=0.4\textwidth]{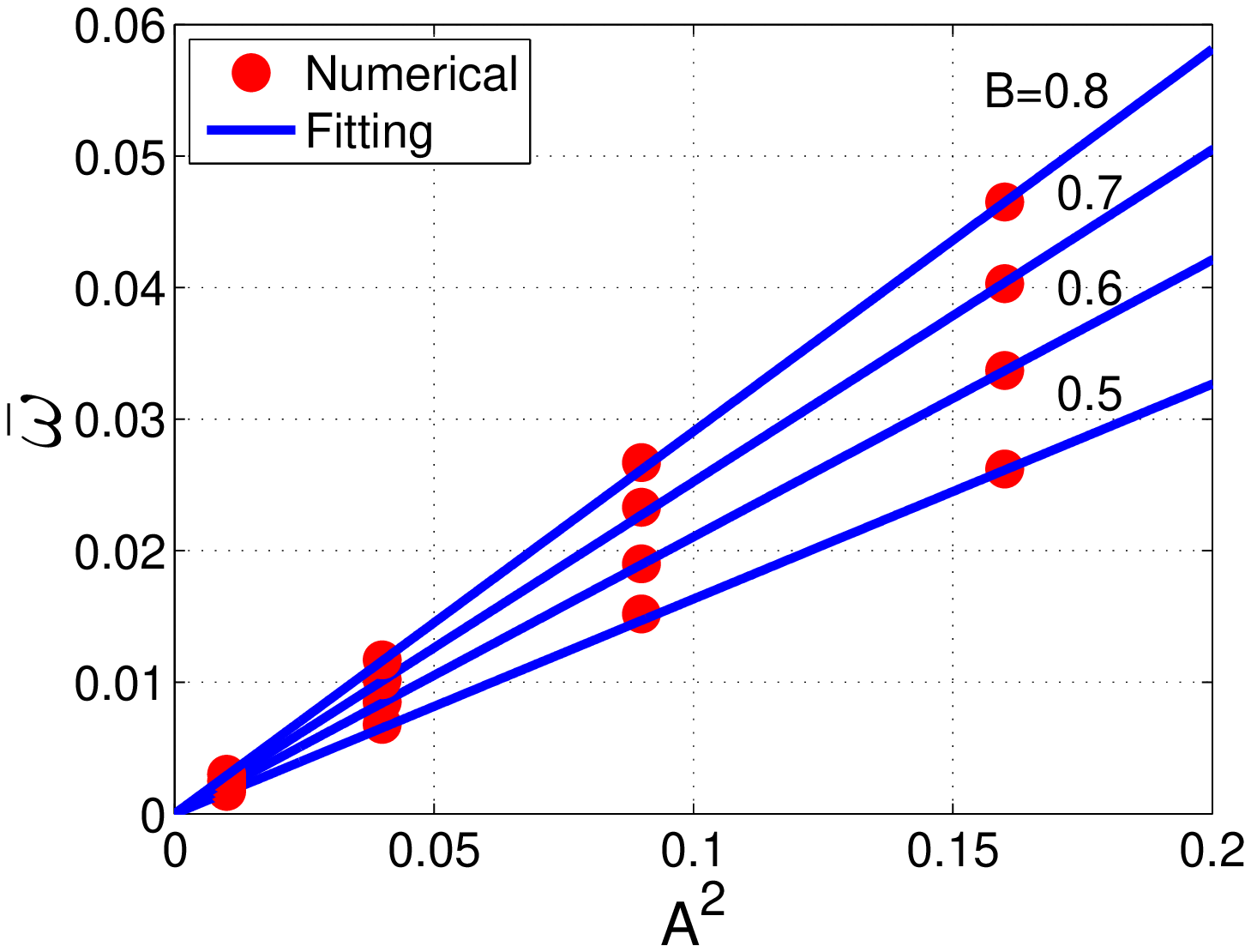} \quad
\includegraphics[width=0.4\textwidth]{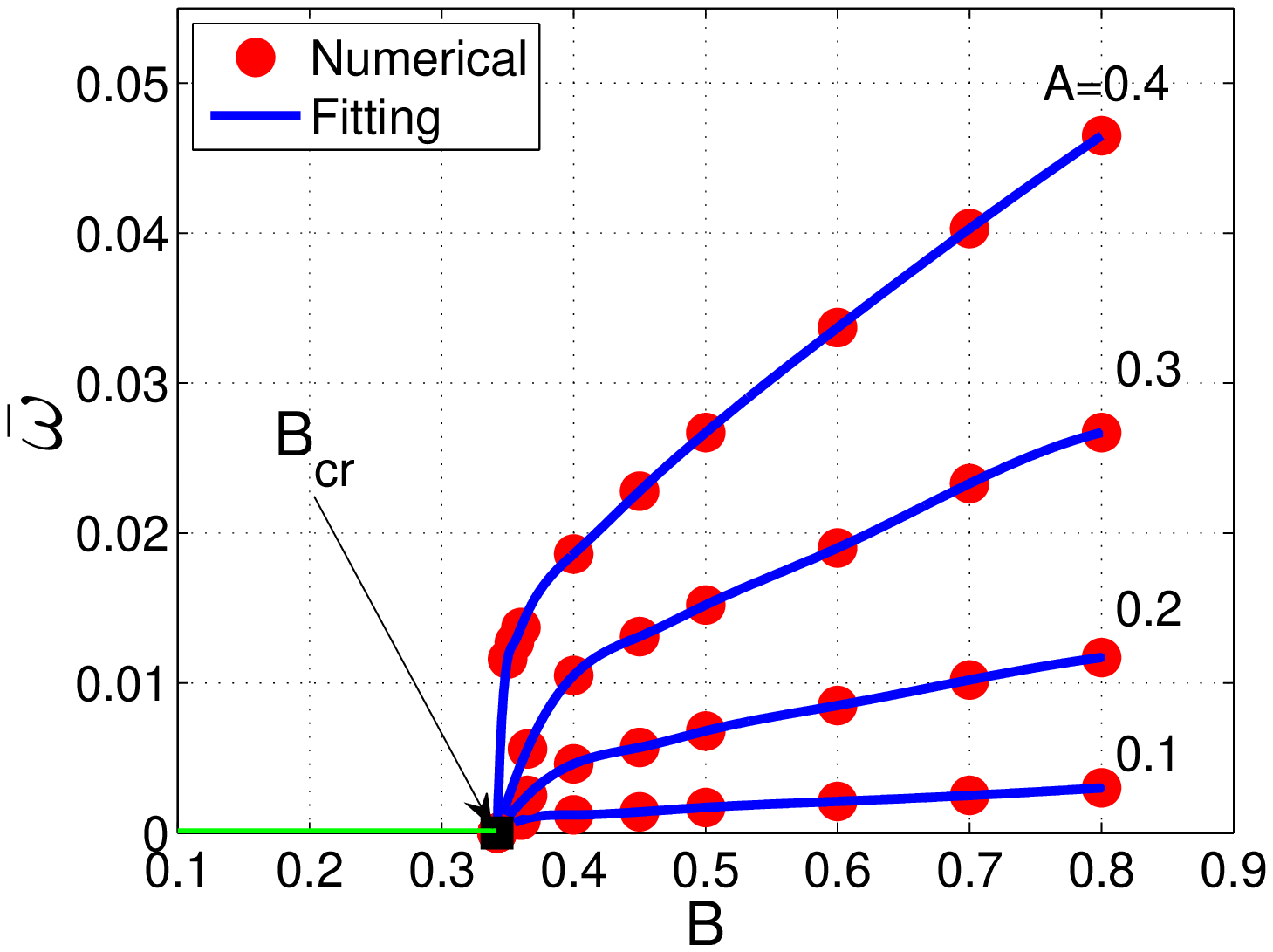}
\caption{Averaged rotation frequency $\bar \omega$ versus amplitude factor $A$ (left panel) and amplitude ratio $B$ (right panel) in a square waveguide.
Red dots are numerical results obtained from Eq.~(\ref{partial}),
blue curves ($B<0.4$) and blue lines ($B>0.4$) are the fitting results to the numerical simulations.
The green line represents the superposed dipoles twist during propagation.
The values of $A$ and $B$ are shown next to the curves.
The black square represents the analytical estimate for $B_{\rm cr}$.}
\label{squarerelations}
\end{figure}

Due to the azimuthon profile deformation, it is not possible to find an analytical expression of $\bar \omega$ as a function of $A$ and $B$ as in the case of circular waveguide (i.e. Eq.~(\ref{analyticalangular})). Therefore, we need to resort to numerical simulations.
In Fig.~\ref{squarerelations} we show the numerically  determined relation between  $\bar \omega$ and  the azimuthon parameters $A$ (left panel) and $B$ (right panel). The threshold-like behavior is evident in the right graph of this figure: the region $B<B_{\rm cr}$, in which the soliton wobbles, is depicted by a green line.

\begin{figure}[htbp]
\centering
\includegraphics[width=0.45\textwidth]{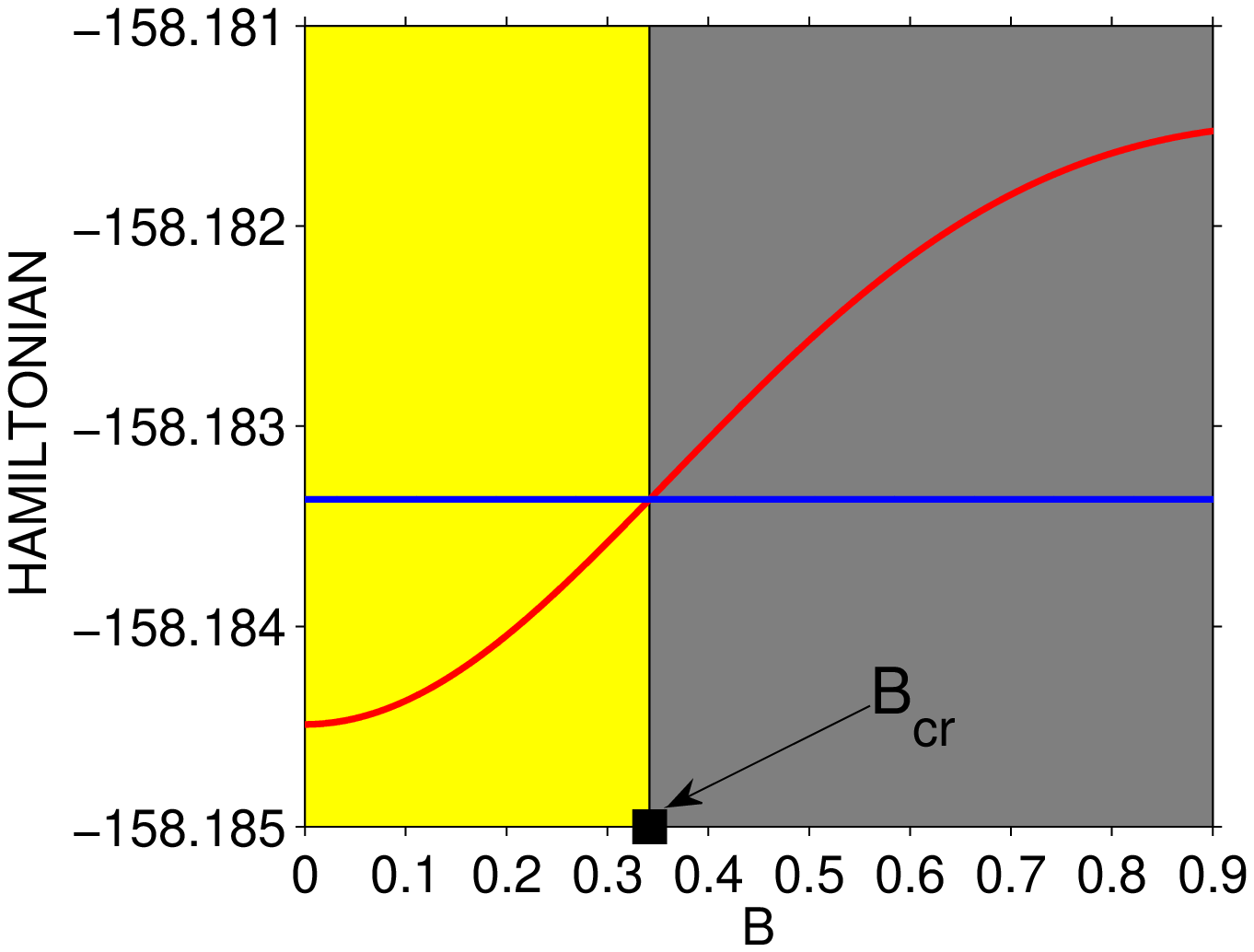} \quad
\includegraphics[width=0.45\textwidth]{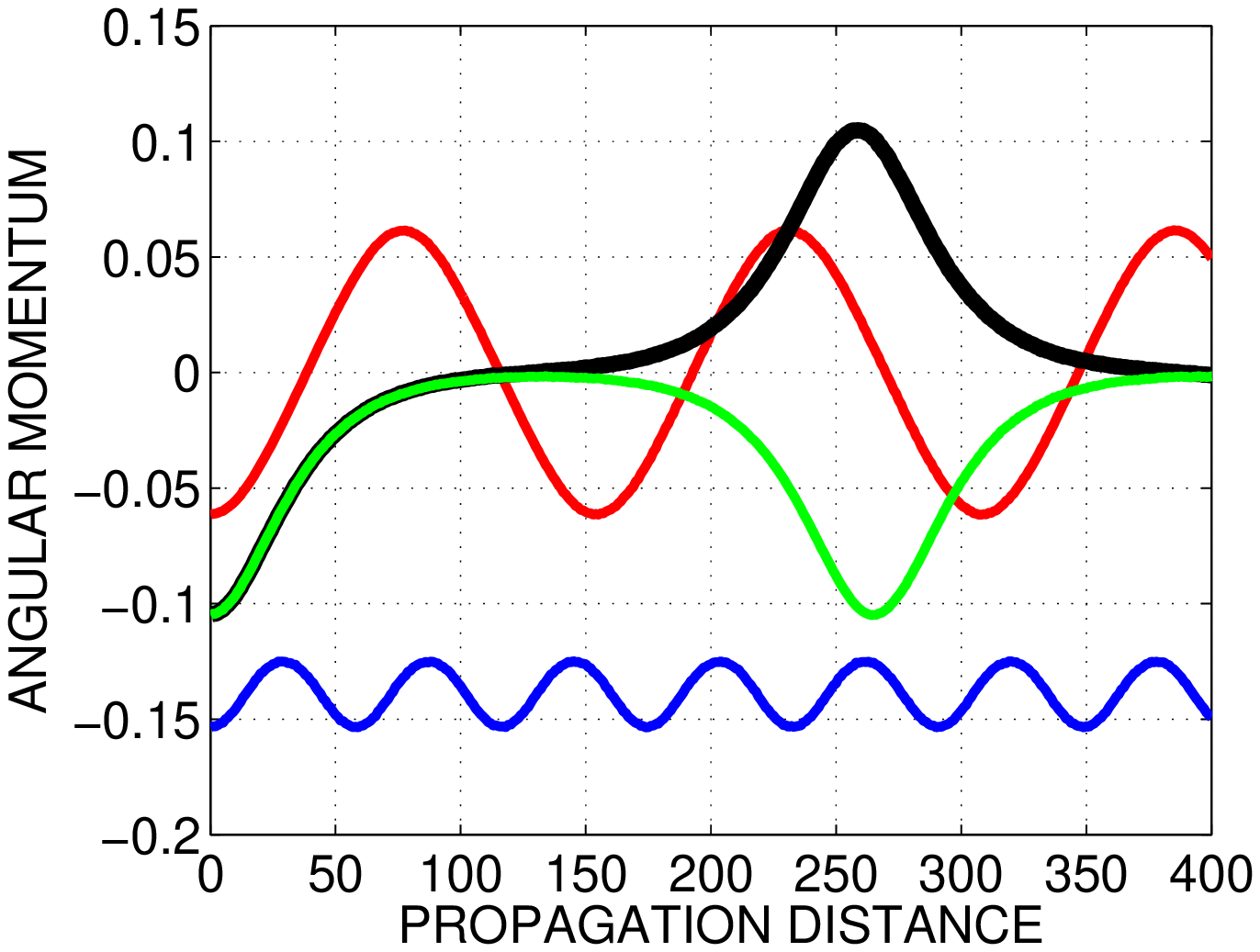}
\caption{
Left panel: Hamiltonian of $A(D_1+iBD_2)$ (red curve) as a function of $B$ and Hamiltonian of $D_3$ (blue line) (for the same power $P$).
The intersection of the two curves defines $B_{\rm cr}$ (depicted by the black square).
Right panel: The angular momenta of the rotating superposed dipoles (corresponding to Fig.~\ref{square_r_evolution}, blue curve),
and the twisting ones (corresponding to Fig.~\ref{square_t_evolution}, red curve).
The other curves depict the cases of $B$ slightly above (black) and bellow (green) $B_{\rm cr}$.
All curves are obtained for $A=0.4$. }
\label{angularmomentumcurve}
\end{figure}

Let us have a closer look at the boundary between 
domains of azimuthon rotation ($B>B_{\rm cr}$) and wobbling ($B<B_{\rm cr}$). We find numerically that the critical value $B_{\rm cr}$, which separates those two domains, depends very weakly on the azimuthon amplitude $A$: we are in weakly nonlinear limit, and may use linear waveguide modes to
elucidate nonlinear propagation properties of the azimuthons.
The appearance of a threshold value $B_{\rm cr}$, above which the azimuthon rotates, can be explained 
by considering the Hamiltonian associated with propagation equation Eq.~(\ref{partial}),
\begin{equation}
\mathcal{H}=\iint \left(|\nabla \psi|^2-\frac{1}{2}|\psi|^4-V|\psi|^2 \right)~\mathrm{d}x ~\mathrm{d}y,
\label{hamiltonian}
\end{equation}
which is a conserved quantity upon propagation.
Depending on their orientation, stationary dipoles ($B=0$) have a slightly different value of $\mathcal{H}$ in nonlinear regime.
If we take our diagonal (w.r.t.\ waveguide cross-section) dipoles $D_1$ and $D_2$ from above,  we can construct a parallel dipole $D_p$ in the following way:
\begin{equation*}
 D_p=\frac{D_1+D_2}{\displaystyle \sqrt{\iint |D_1+D_2|^2 ~\mathrm{d}x ~\mathrm{d}y}}.
\end{equation*}
The intensity distribution of this dipole $D_p$ is very close to that shown in the snapshot of the rotating azimuthon shown in the right panel in Fig.~\ref{square_r_evolution}. One can show using the definition of the Hamiltonian Eq.~(\ref{hamiltonian}) that 
$\mathcal{H}(D_p)>\mathcal{H}(D_1)=\mathcal{H}(D_2)$.
In fact, it turns out that those two dipole orientations (parallel and diagonal) correspond to the extremal values of the Hamiltonian.
Now, if the azimuthon rotates it has to pass through all possible orientations, i.e., its value of $\mathcal{H}$ has to be larger than $\mathcal{H}(D_p)$. We can use this reasoning to estimate the value of $B_{\rm cr}$: 
The left panel of Fig.~\ref{angularmomentumcurve} depicts the dependence of the Hamiltonian of superposed diagonal dipoles as a function of $B$ (red line) for constant power $P$. The blue line represents the value of the Hamiltonian of the parallel dipole $\sim D_p$ at the same power level.  As the red curve monotonically increases with modulation parameter $B$,
the intersection of the two curves gives an estimate of the critical value of the azimuthon modulation $B_{\rm cr}$,
which is shown in the right panel in Fig.~\ref{squarerelations} by a black square.
In other words, in order  to rotate from the diagonal to vertical  position the azimuthon must overcome some kind of energy barrier represented by the difference of the values of  the 
Hamiltonian  in those two basic states. 
As Fig.~\ref{angularmomentumcurve} shows this can be achieved by increasing the value of $B$ in the diagonal state to $B_1=0.342$, which is very close to the value found numerically (see right panel in Fig.~\ref{squarerelations}). Similar phenomena have been identified, for instance, in discrete nonlinear systems where the discreteness introduces the so called Peierls-Nabarro potential which has to be overcome by a soliton to become mobile~\cite{Kivshar_Campbell}.

It should be emphasized that the crucial difference between azimuthon dynamics in  circular and square waveguide originates from the fact  that angular momentum of the beam is conserved in the circular waveguide,
whereas it changes considerably in the square waveguide.
 In the right panel of Fig.~\ref{angularmomentumcurve} we  display the angular momentum of the beam in the square waveguide for four different values of  modulation parameter $B$. 
If $B$ is above the critical value $B_{\rm cr}$, the angular momentum does not change its sign as the blue curve shows.
If  $B<B_{\rm cr}$, the angular momentum changes  periodically its sign  in propagation (red curve),
which indicates the swinging motion of the soliton.   
For values $B$ close to the critical value the variation of the angular momentum  are largest  as shown by the green and black curves. At the same time, the appearance of broad plateaus indicates slow propagation dynamics. 

It is worth mentioning that we also investigated other waveguide symmetries, such as hexagonal waveguides, and found the  behavior of the dipole azimuthons to be qualitatively similar  to  that of the square waveguide, i.e., they either rotate or wobble, depending on modulation depth and initial orientation. However, the difference of the values of the Hamiltonian for the two extremal dipoles is now smaller, resulting in a lower threshold value $B_{\rm cr}$. We believe that this threshold behavior in the propagation dynamics of azimuthons is generic for any non-circular-symmetric waveguide structure which supports degenerated dipole modes.

\section{Rotating higher order localized modes}
\label{higherorder}
In analogy to dipole-azimuthons discussed so far, it is possible to consider dynamics of  higher order
 azimuthons constructed by using pairs of degenerated higher order waveguide modes.
In particular, in a circular waveguide degenerate higher order modes have the form of  quadrupoles, hexapoles, octapoles, decapoles, etc.\, 
and other modes with even number of lobes (optical necklaces).
In a square waveguide, one can identify pairs of degenerate modes in a form of  hexapoles, octapoles,
decapoles, dodecapoles, etc. (optical matrices).
Interestingly, the quadrupoles found in square waveguides are not degenerate 
and hence the propagation dynamics of their corresponding superposed states in weakly nonlinear regime is dominated by linear mode beating.

\begin{figure}[htbp]
\centering
\includegraphics[width=0.45\textwidth]{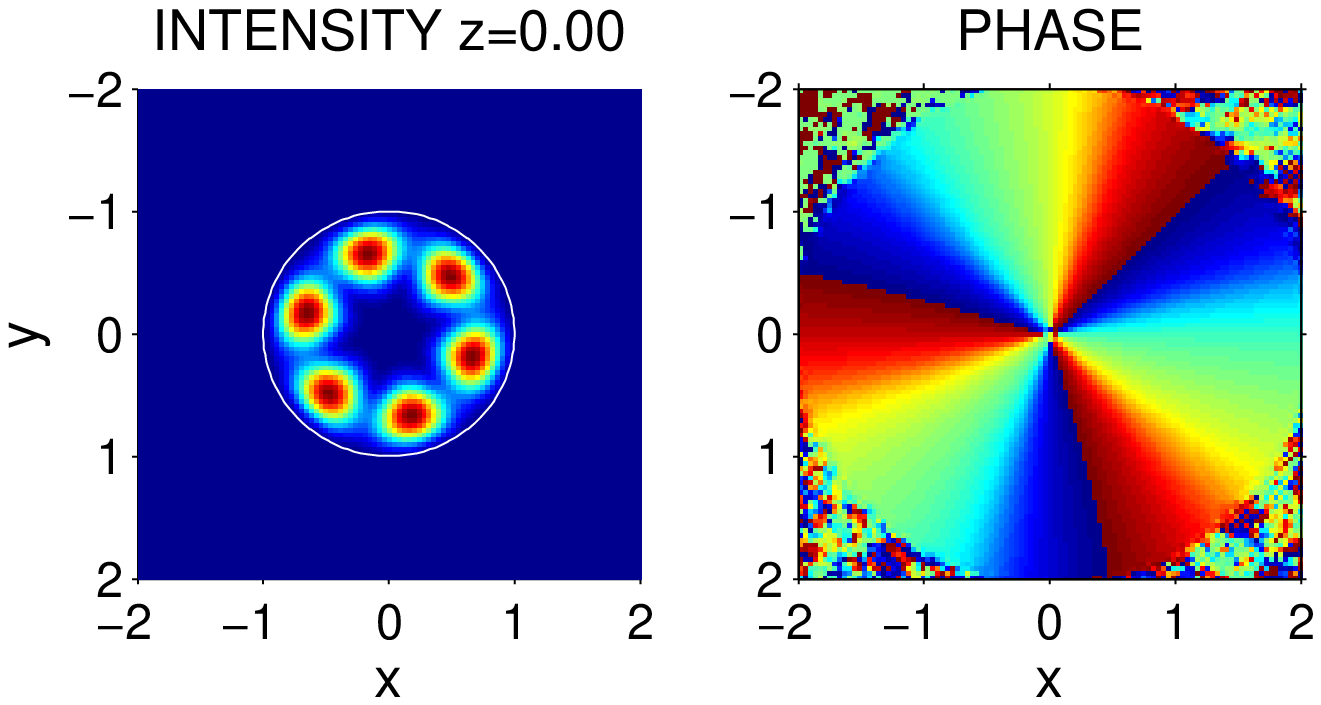} \quad
\includegraphics[width=0.5\textwidth]{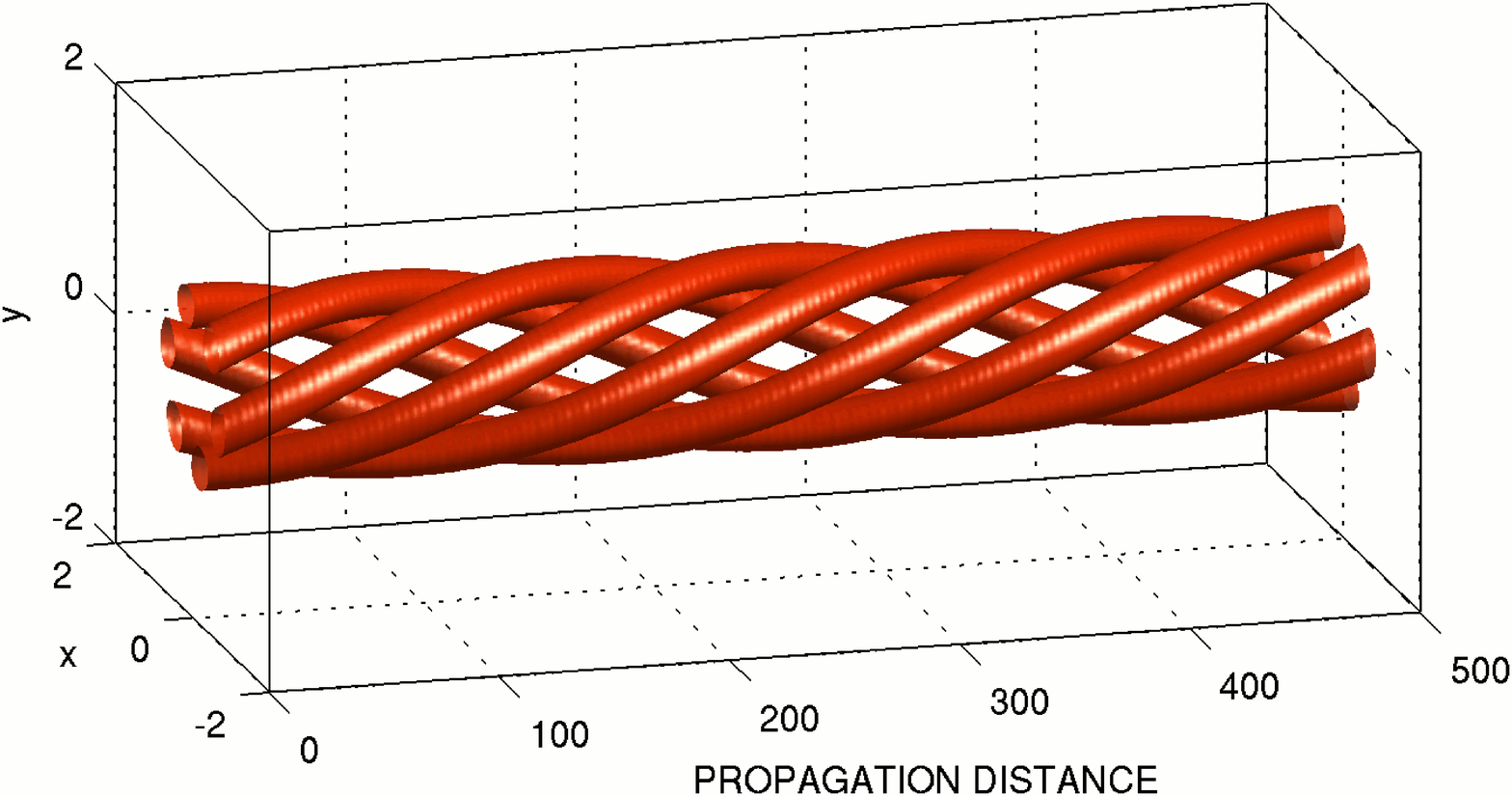}
\caption{The propagation of a hexapole azimuthon with $A=0.4,~B=0.5$ in the circular waveguide.
Left panel shows the original hexapole azimuthon and the corresponding phase;
right panel shows the iso-surface plot of the propagation.}
\label{cquadr_evolution}
\end{figure}

\begin{figure}[htbp]
\centering
\includegraphics[width=0.35\textwidth]{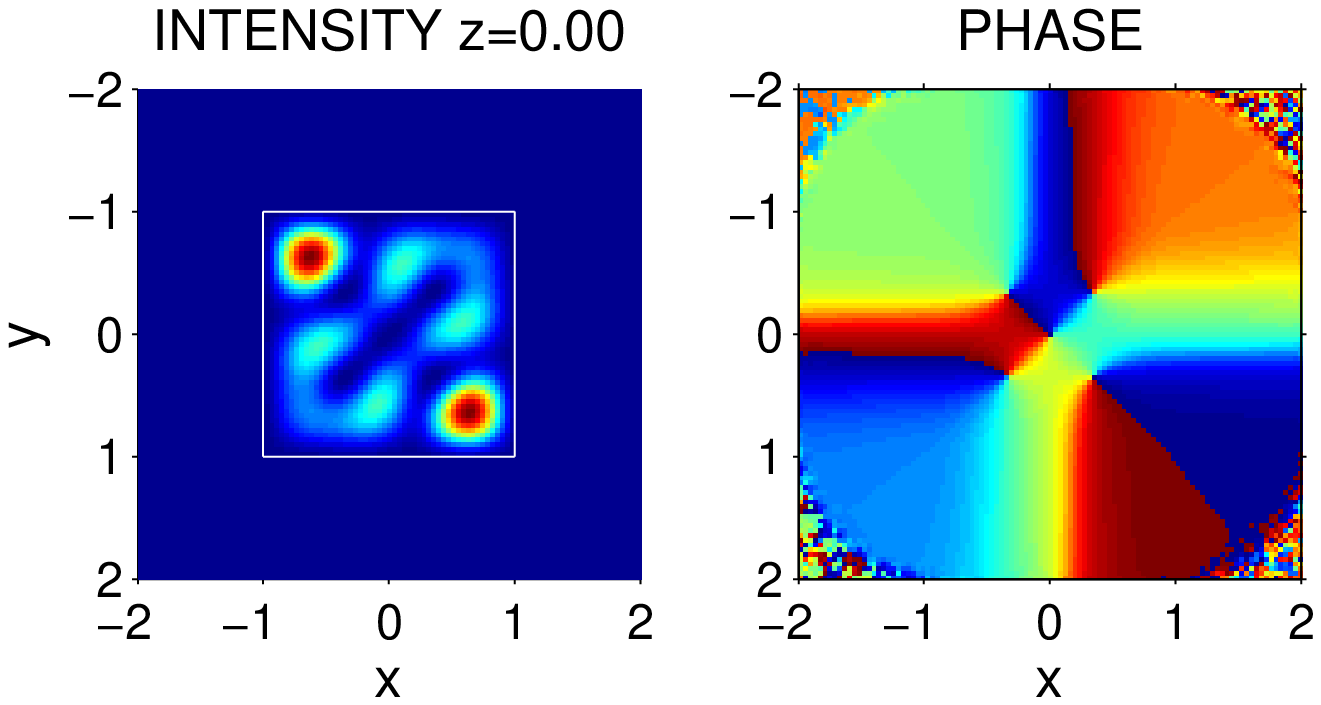} \quad
\includegraphics[width=0.35\textwidth]{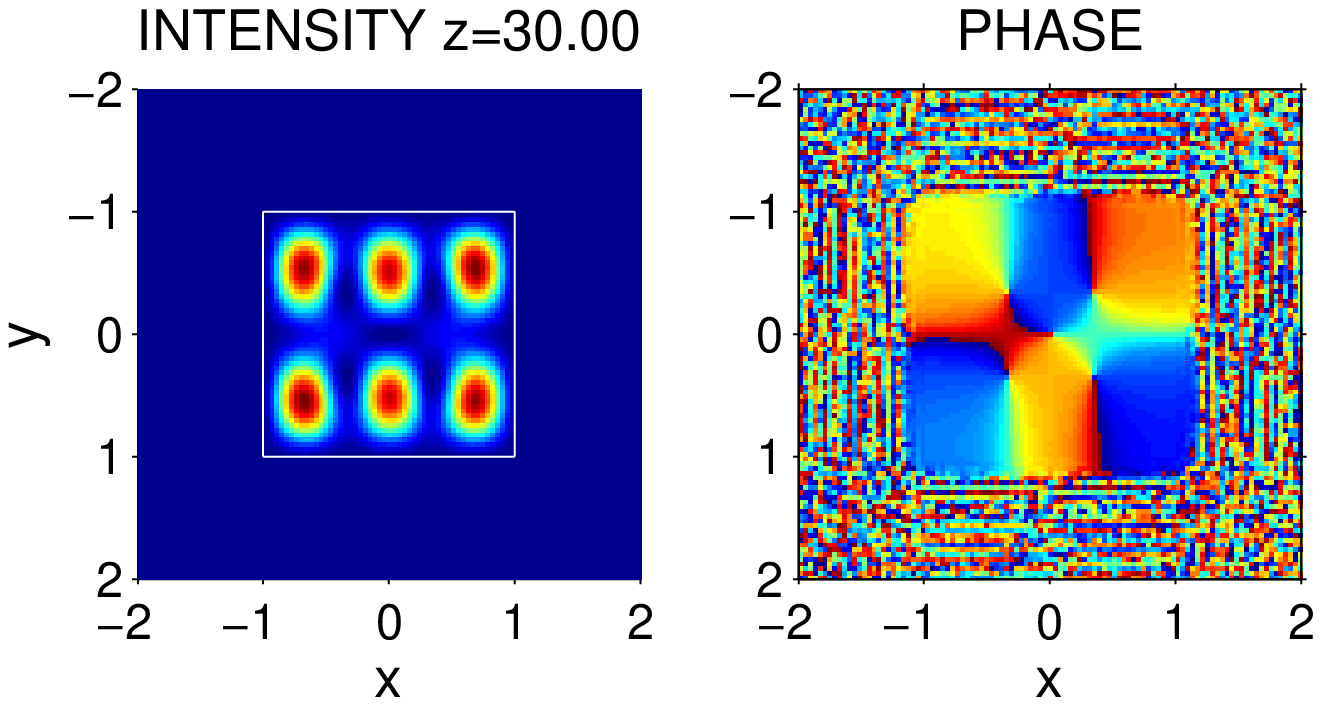} \\
\vspace{0.4cm}
\includegraphics[width=0.35\textwidth]{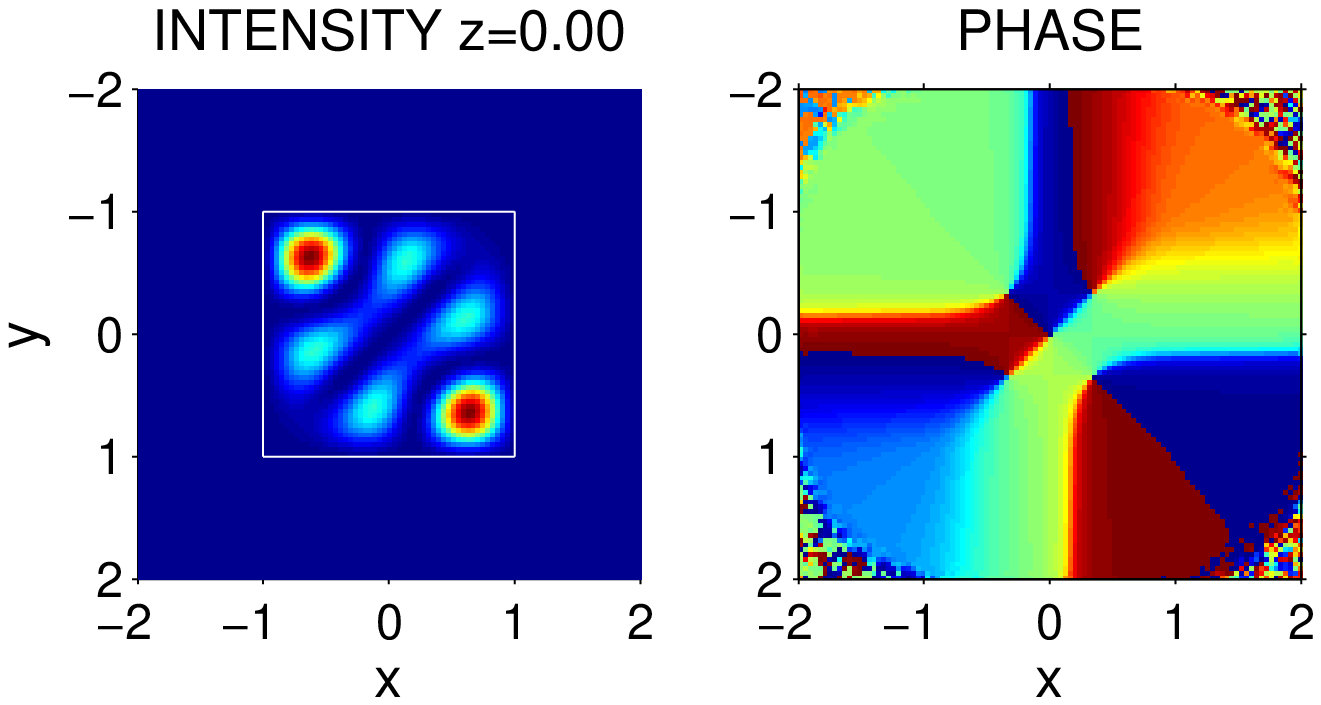} \quad
\includegraphics[width=0.35\textwidth]{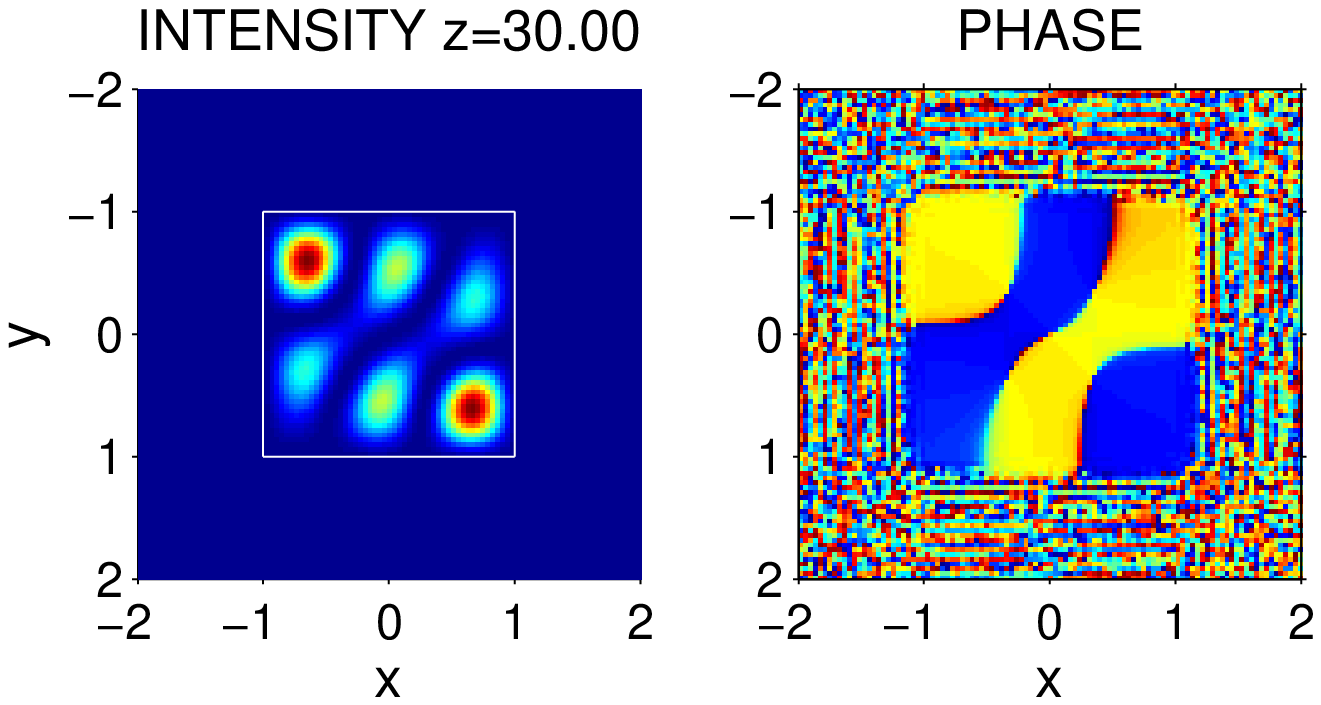} \\
\vspace{0.4cm}
\includegraphics[width=0.35\textwidth]{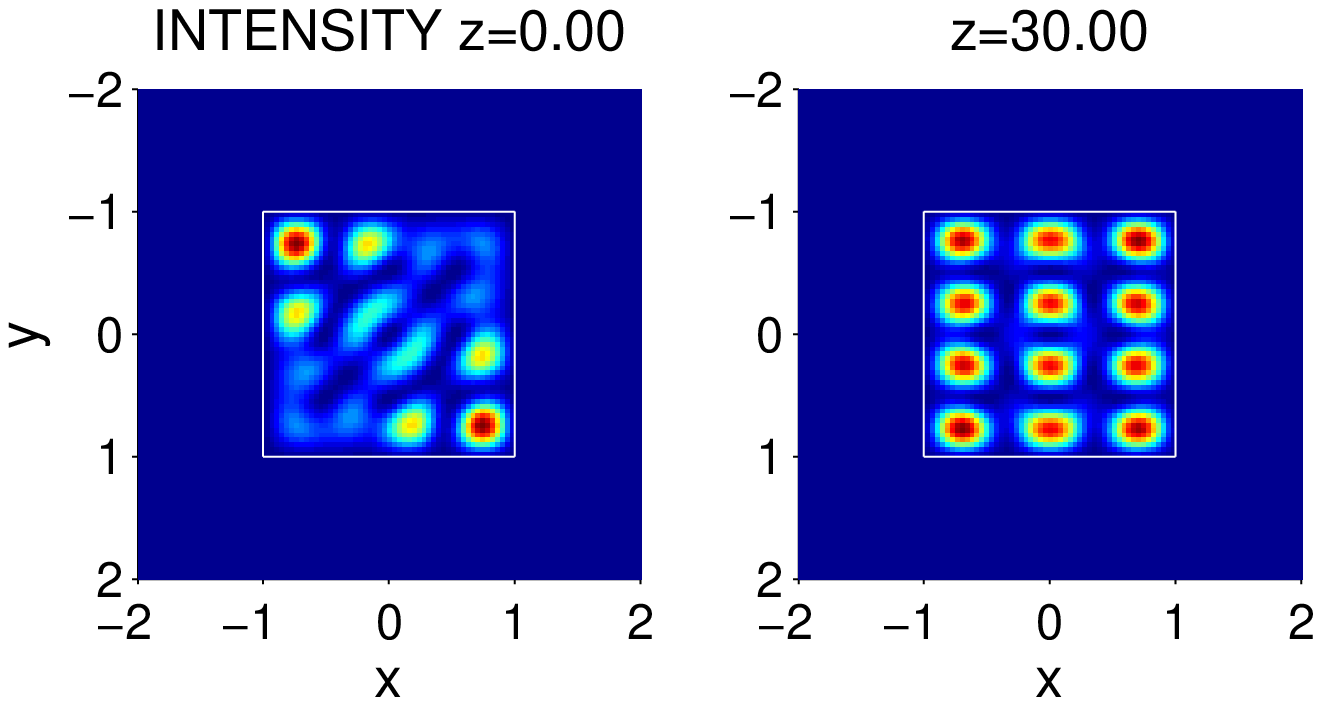} \quad
\includegraphics[width=0.35\textwidth]{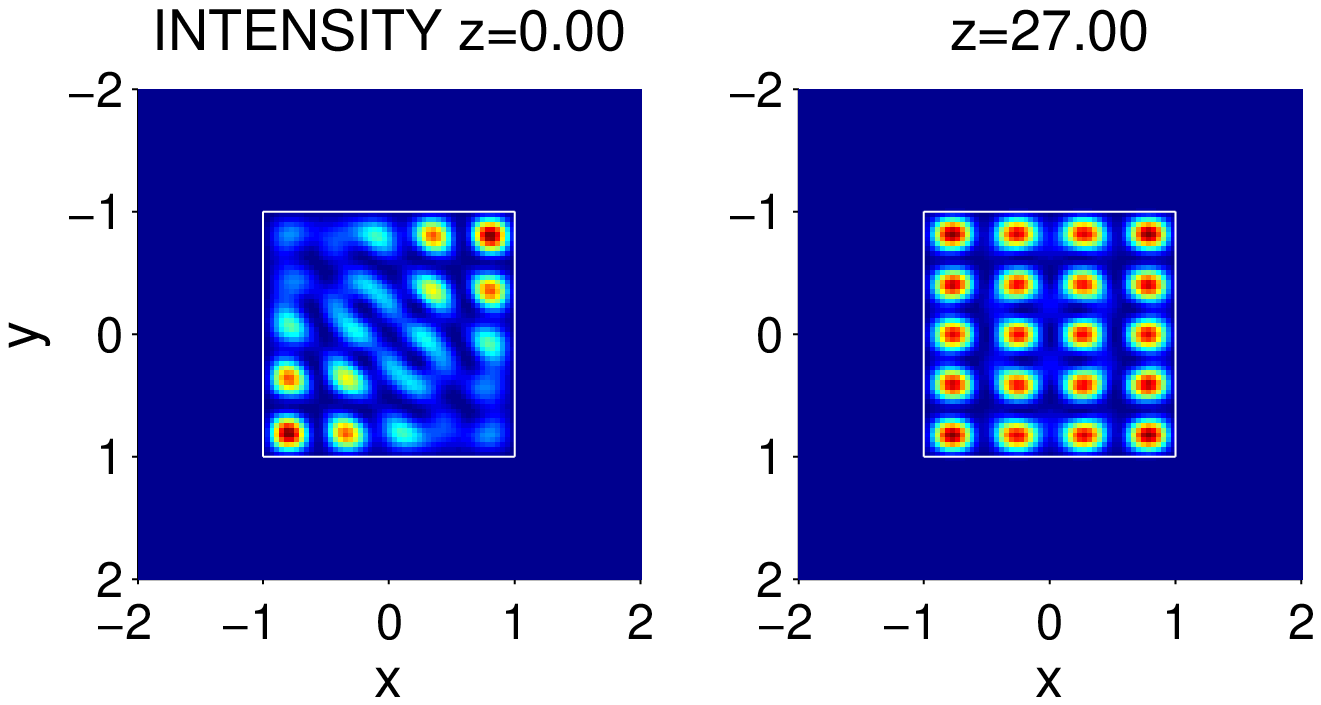}
\caption{The propagation of superposed hexapoles in the square waveguide.
The first row shows a rotating azimuthon with $A=0.4,~B=0.5>B_{\rm cr}$ (Media 1),
and the second row shows a twisting azimuthon with $A=0.4,~B=0.2<B_{\rm cr}$ (Media 2).
The superposed dodecapoles (left two panels),
and superposed icosapoles (right two panels) as well as their deformations are shown in the third row;
in both cases we choose $A=0.4$ and $B=0.5$.}
\label{hexa_evolution}
\end{figure}

Fig.~\ref{cquadr_evolution} displays an example of  hexapole azimuthon in a circular waveguide.
The left panel represents the initial intensity and phase distribution of the azimuthon.  
The 3D surface plot  in the right panel depicts stable and smooth  rotation of the hexapole azimuthon.
As far as the higher order azimuthons of the square waveguide are concerned, they 
 share the dynamical properties of dipole azimuthons  discussed before. 
 As an example,  Fig.~\ref{hexa_evolution} shows the rotating (top row) and swinging (middle row)  hexapoles.
Due to the beam deformation, the hexapole with $B=0.5$ transforms  into a ($2 \times 3$)  matrix while rotating.
The left two panels in the third row of Fig.~\ref{hexa_evolution}  display  the rotating  dodecapole azimuthon  and its deformation  into a ($3 \times 4$) solitonic matrix; the right panels show rotating icosapole azimuthon  and its deformation into a ($4 \times 5$) solitonic  matrix.

\section{Conclusion}
\label{conclusion}
In conclusion, we demonstrated numerically stable propagation of azimuthons, i.e. localized rotating nonlinear waves in weakly nonlinear waveguides.
Depending on the  waveguide profile, different nonlinearity induced propagation dynamics can be observed.
We showed that  azimuthons in circular waveguides  rotate continuously. The analytically predicted dependence of rotation frequency $\omega$ as a function of soliton parameters was found to be in excellent agreement with numerical simulations.
Further on, we discussed propagation of azimuthon-like structures in square waveguides. We showed that their dynamics critically depends on the initial conditions. 
In particular, we found a threshold-like behavior in the propagation dynamics, separating rotating and wobbling solutions.
We showed that this effect is associated with different values of the Hamiltonian for different azimuthon orientations.  
As our analysis relies on physical parameters of actual multi-mode waveguides, our findings may open a relatively easy route to experimental observations of stable azimuthons.

\section{Acknowledgement}
Numerical simulations were partly performed on the SGI
XE Cluster and the Sun Constellation VAYU Cluster of the
Australian Partnership for Advanced Computing (APAC). This
research was supported by the Australian Research Council.


\begin{thebibliography}{10}
\newcommand{\enquote}[1]{``#1''}

\bibitem{Stegeman_science_1999}
G.~I. Stegeman and M.~Segev, \enquote{Optical spatial solitons and their
  interactions: Universality and diversity,} Science \textbf{286}, 1518--1523
  (1999).

\bibitem{desyatnikov_prl_2005}
A.~S. Desyatnikov, A.~A. Sukhorukov, and Y.~S. Kivshar, \enquote{Azimuthons:
  Spatially modulated vortex solitons,} Phys. Rev. Lett. \textbf{95}, 203904
  (2005).

\bibitem{coullet_oc_1989}
P.~Coullet, L.~Gil, and F.~Rocca, \enquote{Optical vortices,} Opt. Commun.
  \textbf{73}, 403--408 (1989).

\bibitem{lashkin_pra_2008}
V.~M. Lashkin, \enquote{Two-dimensional multisolitons and azimuthons in
  {B}ose-{E}instein condensates,} Phys. Rev. A \textbf{77}, 025602 (2008).

\bibitem{buccoliero_ol_2008}
D.~Buccoliero, A.~S. Desyatnikov, W.~Kr{\'o}likowski, and Y.~S. Kivshar,
  \enquote{Spiraling multivortex solitons in nonlocal nonlinear media,} Opt.
  Lett. \textbf{33}, 198--200 (2008).

\bibitem{lopez-aguayo_oe_2006}
S.~Lopez-Aguayo, A.~S. Desyatnikov, and Y.~S. Kivshar, \enquote{Azimuthons in
  nonlocal nonlinear media,} Opt. Express \textbf{14}, 7903--7908 (2006).

\bibitem{buccoliero_prl_2007}
D.~Buccoliero, A.~S. Desyatnikov, W.~Kr{\'o}likowski, and Y.~S. Kivshar,
  \enquote{Laguerre and {H}ermite soliton clusters in nonlocal nonlinear
  media,} Phys. Rev. Lett. \textbf{98}, 053901 (2007).

\bibitem{minovich_oe_2009}
A.~Minovich, D.~N. Neshev, A.~S. Desyatnikov, W.~Kr{\'o}likowski, and Y.~S.
  Kivshar, \enquote{Observation of optical azimuthons,} Opt. Express
  \textbf{17}, 23610--23616 (2009).

\bibitem{kruglov_jmo_1992}
V.~I. Kruglov, Y.~A. Logvin, and V.~M. Volkov, \enquote{The theory of spitral
  laser beams in nonlinear media,} J. Mod. Opt. \textbf{39}, 2277--2291 (1992).

\bibitem{skryabin_pre_1998}
D.~V. Skryabin and W.~J. Firth, \enquote{Dynamics of self-trapped beams with
  phase dislocation in saturable {K}err and quadratic nonlinear media,} Phys.
  Rev. E \textbf{58}, 3916--3930 (1998).

\bibitem{suter_pra_1993}
D.~Suter and T.~Blasberg, \enquote{Stabilization of transverse solitary waves
  by a nonlocal response of the nonlinear medium,} Phys. Rev. A \textbf{48},
  4583--4587 (1993).

\bibitem{bang_pre_2002}
O.~Bang, W.~Kr{\'o}likowski, J.~Wyller, and J.~J. Rasmussen, \enquote{Collapse
  arrest and soliton stabilization in nonlocal nonlinear media,} Phys. Rev. E
  \textbf{66}, 046619 (2002).

\bibitem{krolikowski_job_2004}
W.~Kr{\'o}likowski, O.~Bang, N.~I. Nikolov, D.~Neshev, J.~Wyller, J.~J.
  Rasmussen, and D.~Edmundson, \enquote{Modulational instability, solitons and
  beam propagation in spatially nonlocal nonlinear media,} J. Opt. B: Quantum
  Semiclass. Opt. \textbf{6}, S288--S294 (2004).

\bibitem{briedis_oe_2005}
D.~Briedis, D.~Petersen, D.~Edmundson, W.~Kr{\'o}likowski, and O.~Bang,
  \enquote{Ring vortex solitons in nonlocal nonlinear media,} Opt. Express
  \textbf{13}, 435--443 (2005).

\bibitem{lopez-aguayo_ol_2006}
S.~Lopez-Aguayo, A.~S. Desyatnikov, Y.~S. Kivshar, S.~Skupin,
  W.~Kr{\'o}likowski, and O.~Bang, \enquote{Stable rotating dipole solitons in
  nonlocal optical media,} Opt. Lett. \textbf{31}, 1100--1102 (2006).

\bibitem{stefan_oe_2008}
S.~Skupin, M.~Grech, and W.~Kr{\'o}likowski, \enquote{Rotating soliton
  solutions in nonlocal nonlinear media,} Opt. Express \textbf{16}, 9118--9131
  (2008).

\bibitem{Fabian:oqe:09}
F.~Maucher, D.~Buccoliero, S.~Skupin, M.~Grech, A.~S. Desyatnikov, and
  W.~Krolikowski, \enquote{{Tracking azimuthons in nonlocal nonlinear media},}
  {Opt. Quant. Electron.} \textbf{{41}}, {337--348} ({2009}).

\bibitem{peccianti_ol_2002}
M.~Peccianti, K.~A. Brzdkiewicz, and G.~Assanto, \enquote{Nonlocal spatial
  soliton interactions in nematic liquid crystals,} Opt. Lett. \textbf{27},
  1460--1462 (2002).

\bibitem{conti_prl_2003}
C.~Conti, M.~Peccianti, and G.~Assanto, \enquote{Route to nonlocality and
  observation of accessible solitons,} Phys. Rev. Lett. \textbf{91}, 073901
  (2003).

\bibitem{conti_prl_2004}
C.~Conti, M.~Peccianti, and G.~Assanto, \enquote{Observation of optical spatial
  solitons in a highly nonlocal medium,} Phys. Rev. Lett. \textbf{92}, 113902
  (2004).

\bibitem{nath_pra_2007}
R.~Nath, P.~Pedri, and L.~Santos, \enquote{Soliton-soliton scattering in
  dipolar {B}ose-{E}instein condensates,} Phys. Rev. A \textbf{76}, 013606
  (2007).

\bibitem{lashkin_pra_2009}
V.~M. Lashkin, E.~A. Ostrovskaya, A.~S. Desyatnikov, and Y.~S. Kivshar,
  \enquote{Vector azimuthons in two-component {B}ose-{E}instein condensates,}
  Phys. Rev. A \textbf{80}, 013615 (2009).

\bibitem{litvak_plasmas_1975}
A.~G. Litvak, V.~Mironov, G.~Fraiman, and A.~Yunakovskii, \enquote{Thermal
  self-effect of wave beams in plasma with a nonlocal nonlinearity,} Sov.\ J.\
  Plasma Phys. \textbf{1}, 31--37 (1975).

\bibitem{rotschild_prl_2005}
C.~Rotschild, O.~Cohen, O.~Manela, M.~Segev, and T.~Carmon, \enquote{Solitons
  in nonlinear media with an infinite range of nonlocality: First observation
  of coherent elliptic solitons and of vortex-ring solitons,} Phys. Rev. Lett.
  \textbf{95}, 213904 (2005).

\bibitem{stefan_pre_2004}
S.~Skupin, U.~Peschel, L.~Berg{\'e}, and F.~Lederer, \enquote{Stability of
  weakly nonlinear localized states in attractive potentials,} Phys. Rev. E
  \textbf{70}, 016614 (2004).

\bibitem{boyd_book_2008}
R.~W. Boyd, \emph{Nonlinear Optics} (Academic Press, Amsterdam, 2008), 3rd ed.

\bibitem{kuzuu_ao_1999}
N.~Kuzuu, K.~Yoshida, H.~Yoshida, T.~Kamimura, and N.~Kamisugi,
  \enquote{Laser-induced bulk damage in various types of vitreous silica at
  1064, 532, 355, and 266 nm: Evidence of different damage mechanisms between
  266-nm and longer wavelengths,} Appl. Opt. \textbf{38}, 2510--2515 (1999).

\bibitem{agrawal_book_2009}
G.~P. Agrawal, \emph{Nonlinear Fiber Optics} (Academic Press, Singapore, 2009),
  4th ed.

\bibitem{skryabin_pre_2002}
D.~V. Skryabin, J.~M. McSloy, and W.~J. Firth, \enquote{Stability of spiralling
  solitary waves in {H}amiltonian systems,} Phys. Rev. E \textbf{66}, 055602
  (2002).

\bibitem{fabian_pra_2010}
F.~Maucher, S.~Skupin, M.~Shen, and W.~Kr{\'o}likowski, \enquote{Rotating
  three-dimensional solitons in {B}ose-{E}instein condensates with gravitylike
  attractive nonlocal interaction,} Phys. Rev. A \textbf{81}, 063617 (2010).

\bibitem{Kivshar_Campbell}
Y.~S. Kivshar and D.~K. Campbell, \enquote{{P}eierls-{N}abarro potential
  barrier for highly localized nonlinear modes,} Phys. Rev. E \textbf{48},
  3077--3081 (1993).

\end{thebibliography}
\end{document}